\documentclass[12pt]{article}
\newcommand{\lsim}{\mathrel{\mathop{\kern 0pt \rlap
  {\raise.2ex\hbox{$<$}}}
  \lower.9ex\hbox{\kern-.190em $\sim$}}}
\newcommand{\gsim}{\mathrel{\mathop{\kern 0pt \rlap
  {\raise.2ex\hbox{$>$}}}
  \lower.9ex\hbox{\kern-.190em $\sim$}}}
\usepackage{epsfig}
\usepackage{color}
\usepackage{graphicx}
\usepackage{lineno}
\usepackage{hyperref}
\usepackage{amsmath}
\usepackage{lscape}
\usepackage{booktabs}
\usepackage{array}    
\newcolumntype{C}[1]{>{\centering\arraybackslash}p{#1}} 
\usepackage{multirow}

\begin{document}
\begin{center}
\vspace*{1.0cm}

{\bf{Electron spectral shape of the third-forbidden $\beta$-decay of $^{87}$Rb measured using a Rb$_2$ZrCl$_6$ crystal scintillator}}

\vskip 1.0cm

{\bf P.~Belli$^{a,b}$,
 R.~Bernabei$^{a,b,}$\footnote{Corresponding
author at: Dipartimento di Fisica, Universit\`{a} di Roma ``Tor Vergata'', I-00133 Rome, Italy. E-mail address:
rita.bernabei@roma2.infn.it (R.~Bernabei)},
 F.~Cappella$^{c}$,
 V.~Caracciolo$^{a,b}$,
 R.~Cerulli$^{a,b}$,
 A.~Incicchitti$^{c,d}$,
 A.~Leoncini$^{a,b}$, 
 V.~Merlo$^{a,b}$,
 S.S.~Nagorny$^{e,g}$,
 V.V.~Nahorna$^{f}$,
 S.~Nisi$^{g}$,
 P.~Wang$^{f}$,
 J. Suhonen$^{h,i}$,
 M. Ramalho$^{j}$,
 J. Kostensalo$^{l}$
 }

\vskip 0.3cm

$^{a}${\it Dipartimento di Fisica, Universit\`{a} di Roma ``Tor Vergata'', I-00133 Rome, Italy}

$^{b}${\it  INFN, sezione di Roma ``Tor Vergata'', I-00133 Rome, Italy}

$^{c}${\it  INFN, sezione di Roma, I-00185 Rome, Italy}

$^{d}${\it  Dipartimento di Fisica, Universit\`{a} di Roma ``La Sapienza'', I-00185 Rome, Italy}

$^{e}${\it Gran Sasso Science Institute, L'Aquila I-67100, Italy}

$^{f}${\it Department of Chemistry, Queen's University, Kingston, ON, K7L 3N6, Canada}

$^{g}${\it INFN Laboratori Nazionali del Gran Sasso, 67100 Assergi (AQ), Italy}

$^{h}${\it Department of Physics, University of Jyv\"askyl\"a, P.O. Box 35, FI-40014, Jyv\"askyl\"a, Finland}

$^{i}${\it International Centre for Advanced Training and Research in Physics (CIFRA), P.O. Box MG12, 077125 Bucharest-M\u{a}gurele, Romania}

$^{j}${\it School of Physics, Engineering and Technology, University of York, Heslington, York YO10 5DD, United Kingdom}

$^{j}${\it Natural Resources Institute Finland, Yliopistokatu 6B, FI-80100 Joensuu, Finland}

\end{center}

\vskip 0.5cm
\begin{abstract}
In recent years, interest in experimental studies of $\beta$-decay electron spectra -- often referred to as 
$\beta$ spectra -- has been growing. This is particularly true for $\beta$ transitions where the electron 
spectra are sensitive to the effective value of the weak axial coupling, $g_{\rm A}$. Such measurements serve 
as important benchmarks for nuclear physics calculations and can also be used to characterize background 
in astroparticle physics experiments. In this work, a dedicated experiment has been carried 
out to investigate the spectral shape of the third-forbidden $^{87}$Rb $\beta$-decays, with the goal of estimating the effective $g_{\rm A}$ 
value for this transition and of deriving the T$_{1/2}$ value. This was done by comparing the experimental spectral shape with 
the estimates from various phenomenological models.
The $^{87}$Rb source was embedded directly within the detector material of
a new Rb$_2$ZrCl$_6$ crystal scintillator; the data taking was performed deep underground at Gran 
Sasso National Laboratory. 
The obtained experimental half-life value for the studied process is T$_{1/2} = 5.08(13) \times$ 10$^{10}$ yr; while a $g_{\rm A}$ value in the range 0.4 to 0.6 
is obtained when accounting for uncertainties and depending on the model adopted as discussed in detail in the text. 
\end{abstract}

\section{Introduction}

The nucleus $^{87}$Rb disintegrates to $^{87}$Sr through $\beta^-$ decay by emitting electrons and electron antineutrinos.
The emitted $\beta$ electrons constitute the 
$\beta$-electron spectral shape, and studies of this shape have become a popular subject of study, e.g. in connection with nuclear reactors and the anomalies in their antineutrino fluxes. The $\beta$ electrons can also be severe background in rare-events experiments searching for beyond-the-standard-model physics, like in those trying to measure rare $\beta$-decays,  neutrinoless double beta decays and dark matter particles. 
In all these cases, measurements of the associated $\beta$ spectral shapes are needed.
These spectral shapes are also an incentive to experiments trying to pin down the effective value of the weak axial coupling $g_{\rm A}$, relevant for the sensitivity estimation of the present and future beyond-the-standard-model experiments, \cite{Eng2017,Eji2019,Ago2023}.

The $\beta$ electrons are emitted in nuclear $\beta^-$-decays, which can be allowed Fermi and Gamow--Teller decays (the emitted lepton pair has no orbital angular momentum, see \cite{Suh2007}) or forbidden decays where the emitted lepton pair has a nonzero orbital angular momentum, see \cite{Beh1982}. 
Forbidden $\beta$ transitions can be classified into unique and non-unique ones, the former having a universal $\beta$ spectral shape \cite{Suh2007}, and the latter having a $\beta$ spectral shape intimately connected to the wave functions of the initial and final states of the transition in the form of the many nuclear matrix elements (NME) contributing to the 
$\beta$-decay amplitude \cite{Beh1982,Haa2016,Haa2017}.

In addition to the many NMEs, the (partial) half-life of a forbidden non-unique $\beta$ transition depends on the values of the weak vector and axial-vector couplings, $g_{\rm V}$ and $g_{\rm A}$, respectively \cite{Beh1982,Haa2016,Haa2017}. The CVC (conserved vector current) hypothesis confines $g_{\rm V}=1.0$ whereas the PCAC (partially conserved axial-vector current) hypothesis allows an effective (quenched) value of $g_{\rm A}^{\rm eff}$ in finite nuclei \cite{Suh2017a}, its bare-nucleon value $g_{\rm A}=1.27$ resulting from experimental studies of the decay of an isolated neutron. The quenching of the effective value of $g_{\rm A}$ has recently been studied extensively in \cite{Eji2019,Suh2017a,Suh2019}, and in particular its effects on the neutrinoless double beta decay have been addressed in \cite{Eng2017,Ago2023,Suh2017b}. Analyses of the quenched value of $g_{\rm A}$ exploiting computed and measured $\beta$-spectral shapes of individual $\beta^-$ transitions have recently been performed for the fourth-forbidden non-unique $\beta$-decays of $^{113}$Cd \cite{Bod2020,Kos2021,Kos2023,Ban2024} and $^{115}$In \cite{Led2022,Pag2024,Kos2024}. There is also a recent measurement of the $\beta$ spectrum of the second-forbidden non-unique $\beta$-decay of $^{99}$Tc \cite{Pau2023} leading to an enhancement of $g_{\rm A}$ when the CVC value of $g_{\rm V}$ is assumed. Contrary to this, the works \cite{Ram2024,Ram2024b} would point to a quenched value of $g_{\rm A}$.

The present work reports the recent study of the spectral shape of the third-forbidden non-unique ground-state-to-ground-state $\beta$ transition $^{87}\textrm{Rb}(3/2^-)\to\,^{87}\textrm{Sr}(9/2^+)$ (released energy Q$_{\beta}=282.275(6)$ keV \cite{AME2020}), with a 100\% branching ratio
and a current recommended half-life value of T$_{1/2} = 4.9650(40) \times 10^{10}$ yr \cite{DDEP2025}.
The $^{87}$Rb atoms are present in the natural Rubidium with a rather large natural abundance ($\delta = 0.2783(2)$ \cite{Meija2016}).

For this purpose a new Rb$_2$ZrCl$_6$ crystal scintillator has been developed and produced 
following the previous successful
realizations of the Cs$_2$XCl$_6$ crystal family scintillators, originally
proposed to study rare processes in X=Hf \cite{Hf1,Hf2} and X=Zr
\cite{Zr1,Zr2} isotopes\footnote{The new Rb$_2$ZrCl$_6$ crystal scintillator and the 
Cs$_2$XCl$_6$ crystal family scintillators have been proposed, developed and produced under the supervision of Serge Nagorny.}.
The use of the Rb$_2$ZrCl$_6$ crystal scintillator allows for optimal application of the ``source = detector" approach,
which guarantees about 100\% efficiency for the detection of the emitted electrons in the $^{87}$Rb $\beta$-decay.
Details on the production process of the crystal and its characterization are given in Section \ref{sec_0} together with
the description of the experimental setup, the calibrations, and the data taking.
The study of the experimental $\beta$ spectrum and its unfolding procedure is described in Section \ref{sec_beta}.
The following Sections are dedicated to describe calculations of the $\beta$ spectral shapes of $^{87}$Rb for different values of $g_{\rm A}$ and different nuclear models
and to compare them with the experimental one, in order to derive the effective value of $g_{\rm A}$.

\section{Experimental setup and data taking}
\label{sec_0}

The development of the Rb$_2$ZrCl$_6$ (RZC) crystal scintillator has 
played a crucial role in the investigations performed in this work, 
enabling the effective ``source = detector'' approach. In this approach, 
the source of decaying $^{87}$Rb nuclei is embedded directly within 
the detector material itself, ensuring both high detection efficiency 
and excellent energy resolution. Thus, to provide further context,
a brief description of the production process for this crystal is 
provided in next subsection. Then, the following subsection gives 
the description of the measurements and of the obtained experimental results.

\subsection{The Rb$_2$ZrCl$_6$ crystal production and its purity}

Anhydrous RbCl (99.8\%) and ZrCl$_4$ (99.9\%) powders were used as starting materials.  
The ZrCl$_4$ powder was first subjected to a single-stage sublimation under dynamic vacuum to purify it. 
Next, RbCl and purified ZrCl$_4$ powders were mixed in a stoichiometric ratio and loaded into a tapered quartz ampoule. 
Then, the mixture was dried at 120$^o$C for 24 hours under vacuum to remove moisture before sealing.
The Rb$_2$ZrCl$_6$ crystal was grown using the vertical Bridgman-Stockbarger technique in a two-zone furnace. 
The sealed ampoule containing the reagents was gradually heated at a rate of 0.3$^o$C/min and held at 
850$^o$C for 24 hours to synthesize the stoichiometric compound and ensure melt homogeneity. 
Crystal growth was initiated with a pulling rate of 24 mm/day and a temperature gradient of 
30$^o$C/cm in the crystallization zone. The growth rate was then reduced to 12 mm/day with a temperature 
gradient of 15$^o$C/cm for the second stage. 
From the obtained boule, a
crystal of Rb$_2$ZrCl$_6$ with a diameter of 21 mm 
and a length of approximately 40 mm was obtained. 
A crystal sample weighting about 0.49 g was extracted and used, as even such small amount is sufficient given the induced counting rate.
In fact, this allows for the detection of $\simeq 10^{7}$ $\beta$-decay events of $^{87}$Rb 
within one day of measurements.
 An additional advantage of using such a small mass sample is the significant
reduction of unintended background events, enhancing the precision of the measurements.

The concentration of some important chemical contaminations in raw materials and RZC crystal sample was determined via 
Inductively Coupled Plasma Mass-Spectrometry (ICP-MS). 
The analysis was done in a semi-quantitative mode, calibrating the instrument with a single standard solution containing 10 ppb 
of Li, Y, Ce, and Tl. The uncertainty is about 25\% of the given concentration values. 
About 50 mg of each sample were placed in a plastic vial, 
to which 5 ml of ultra-pure water and 0.5 ml of nitric acid were added; the vial was then placed in an ultrasonic bath at 60$^{o}$C temperature
until complete decomposition of the sample.
The vials with the samples dissolved in this manner were prepared for ICP-MS analysis 
by adding ultra-pure water up to 10 ml of total volume, i.e., with a dilution factor of about 2000.
The results are given in Table \ref{conc}. 

\begin{table}[!hbt]
\begin{center}
\caption{Concentration of some important chemical contamination in the Rb$_2$ZrCl$_6$ crystal and in its starting materials as measured by ICP-MS analysis. 
The uncertainty is about 25\% of the given concentration values expressed in ppb.}
\begin{tabular}{|c|c|c|c|}
  \hline
  Element	& RbCl	& ZrCl$_4$	& Rb$_2$ZrCl$_6$ \\
 \hline
K	& 33\,400	& 16\,900	& 17\,000 \\
Cs	& 600\,000 &	4\,500	 & 300\,000 \\
Hf	& $<$ 5	& 16\,700	& 8\,900 \\
Pb	& 950	& 250	& 225 \\
Th	& $<$ 0.2	& 140	& 4  \\
U &	0.2 &	1\,500	 & 54 \\
  \hline
\end{tabular}
\label{conc}
\end{center}
\end{table}

\subsection{Experimental setup}
\label{sec_exp}
\begin{figure}[!htp]
\centering
\includegraphics[scale=0.19]{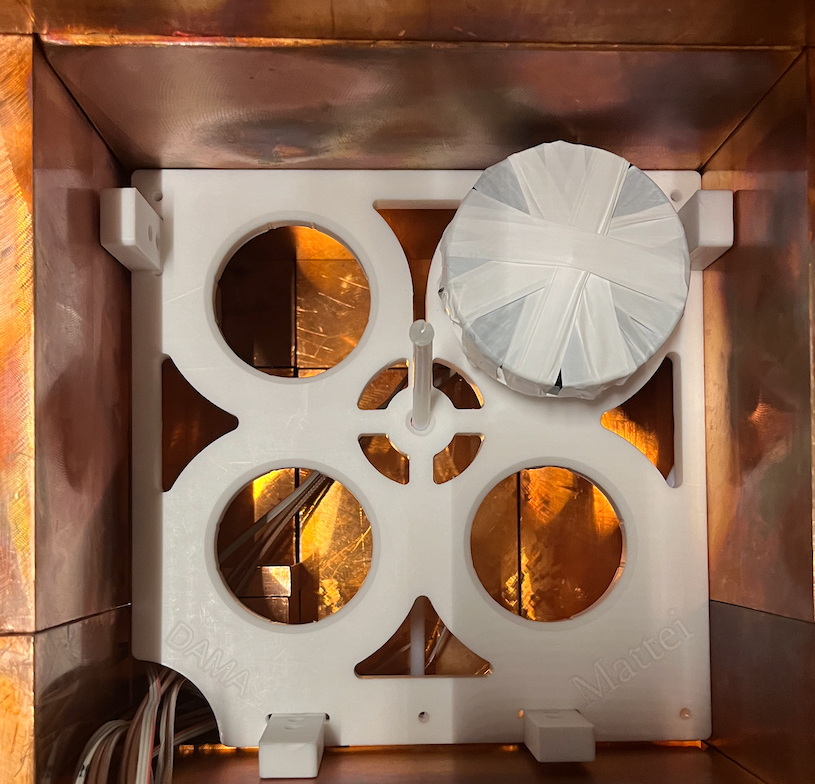} 
\hspace{0.3mm}
\includegraphics[scale=0.173]{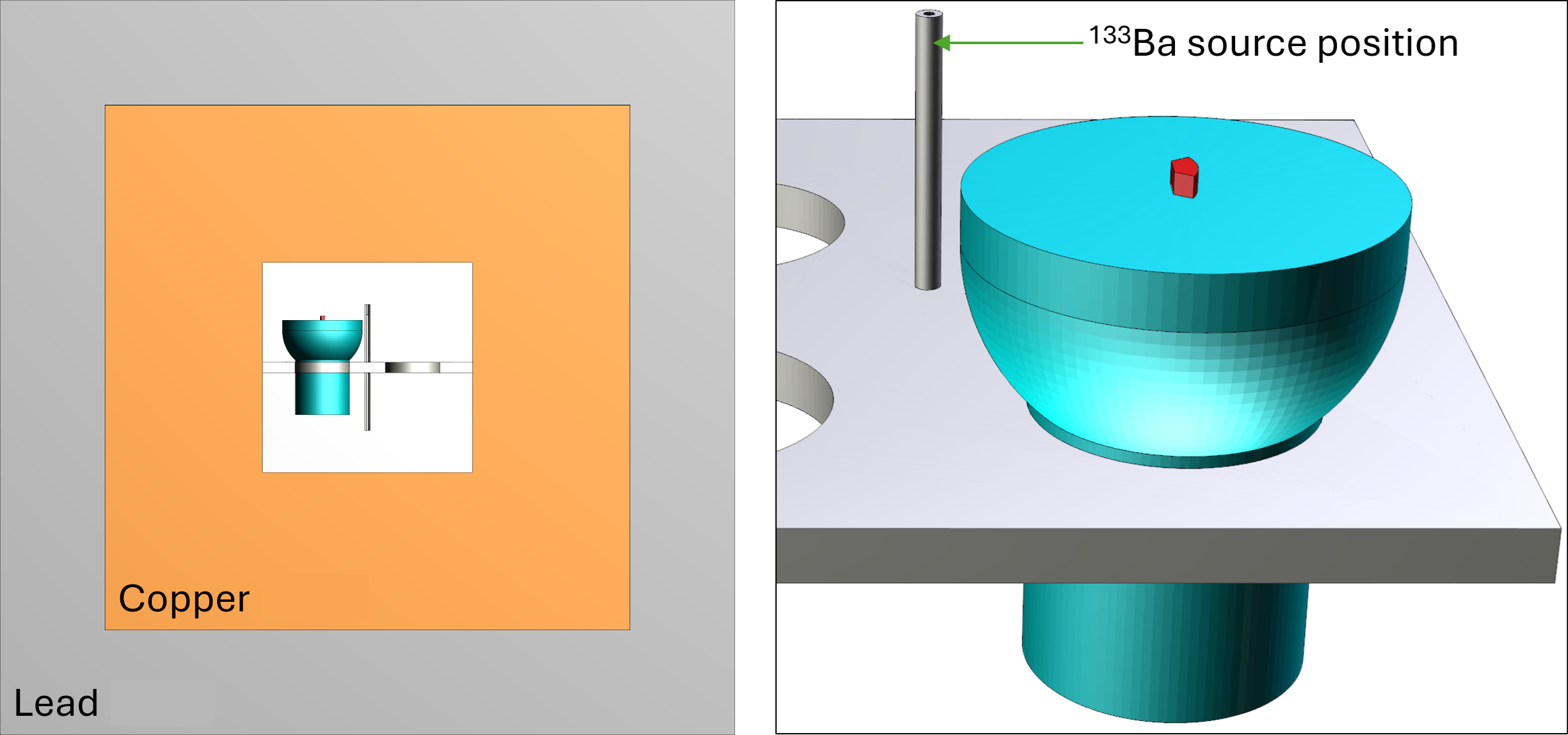} 
 \caption{Photo and a sketch of the experimental setup. The position of the $^{133}$Ba calibration source in form of wire can also be identified.
The optical coupling between the RZC crystal scintillator and the  3-inch ultra-low-background PMT was guaranteed by optical couplant. The crystal and the PMT 
were secured by PTFE tape as shown in the photo (not in the sketch).}
 \label{fig_schema}
\end{figure}

To identify the nuclear process under investigation, determine its spectral shape, and derive the value of $g_{\rm A}$, 
the detector was assembled by optically coupling the RZC crystal scintillator to a 3-inch ultra-low-background photo-multiplier,
PMT 
(HAMAMATSU R6233MOD \cite{contamin}), by means of optical couplant.
The RZC crystal was wrapped by multilayer of 
polytetrafluoroethylene (PTFE)
reflective tape with a thickness of about 1.5 mm and secured by PTFE tape to PMT, as shown in Fig. \ref{fig_schema}. 
Although this measurement is not a low-background experiment, it was conducted 
within the DAMA/CRYS ultra-low-background facility \cite{damacrys}, located deep underground at Gran Sasso laboratory (LNGS) in the
by-pass between hall A and hall B. The passive shield was made of an innermost layer of 15 cm of
Oxygen-Free High Conductivity  (OFHC)
Cu, followed by 20 cm of low-radioactive Pb and by 5 cm of 
High-Density (HD) polyethylene as external layer.
A photo and a sketch of the experimental setup are reported in Fig. \ref{fig_schema}.
Specifically, the RZC crystal scintillator was realized from the 0.49 g sample from the grown boule as mentioned above;
after the surface treatments, it has a mass $m = 0.4647(1)$ g with irregular shape and with maximum sizes approximately of $9 \times 6 \times 4$ mm$^3$.
Taking into account that 300 ppm of Cs and 17 ppm of K contaminants (see Table \ref{conc}) contribute to the total mass on the order of 0.03\%, 
we obtained a minor correction to the number of $^{87}$Rb nuclei in 
the sample $N_{87} = 3.279(2) \times 10^{20}$ (see Ref. \cite{tabref}).  

\subsection{Scintillation pulses}
\label{sec_pulse}

\begin{figure}[!ht]
\centering
\includegraphics[scale=0.25]{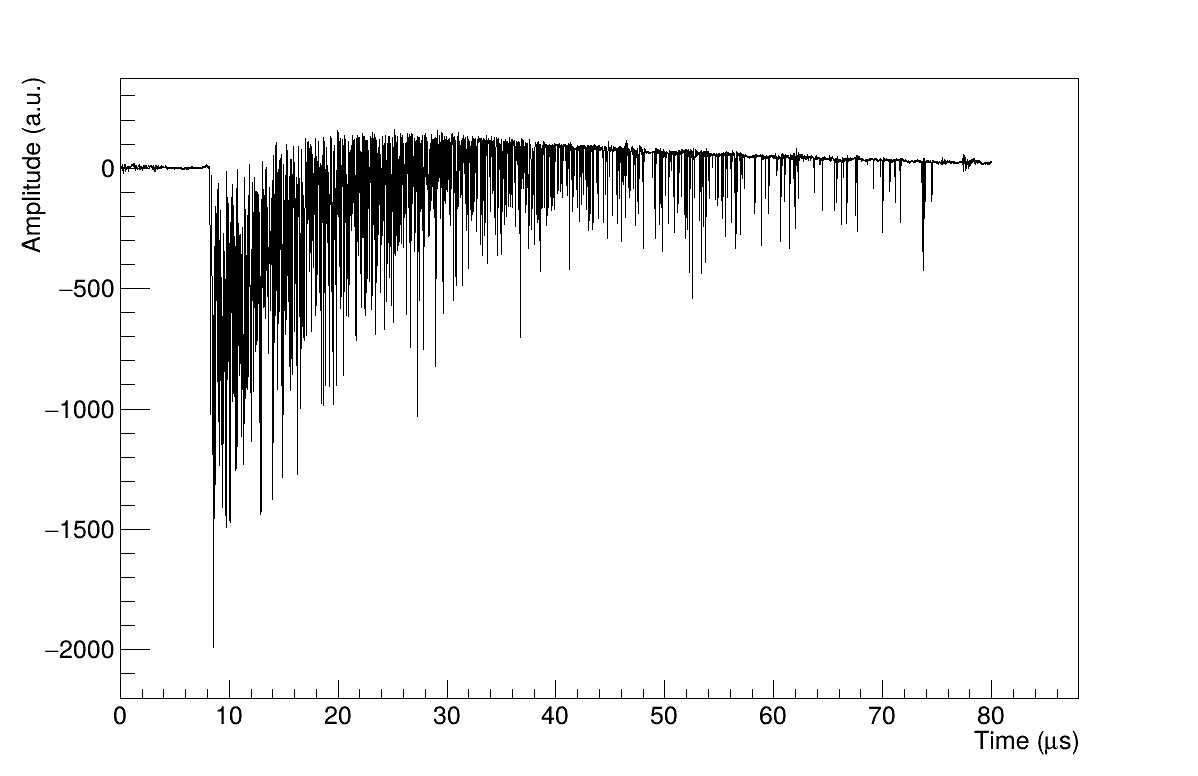} 
\includegraphics[scale=0.25]{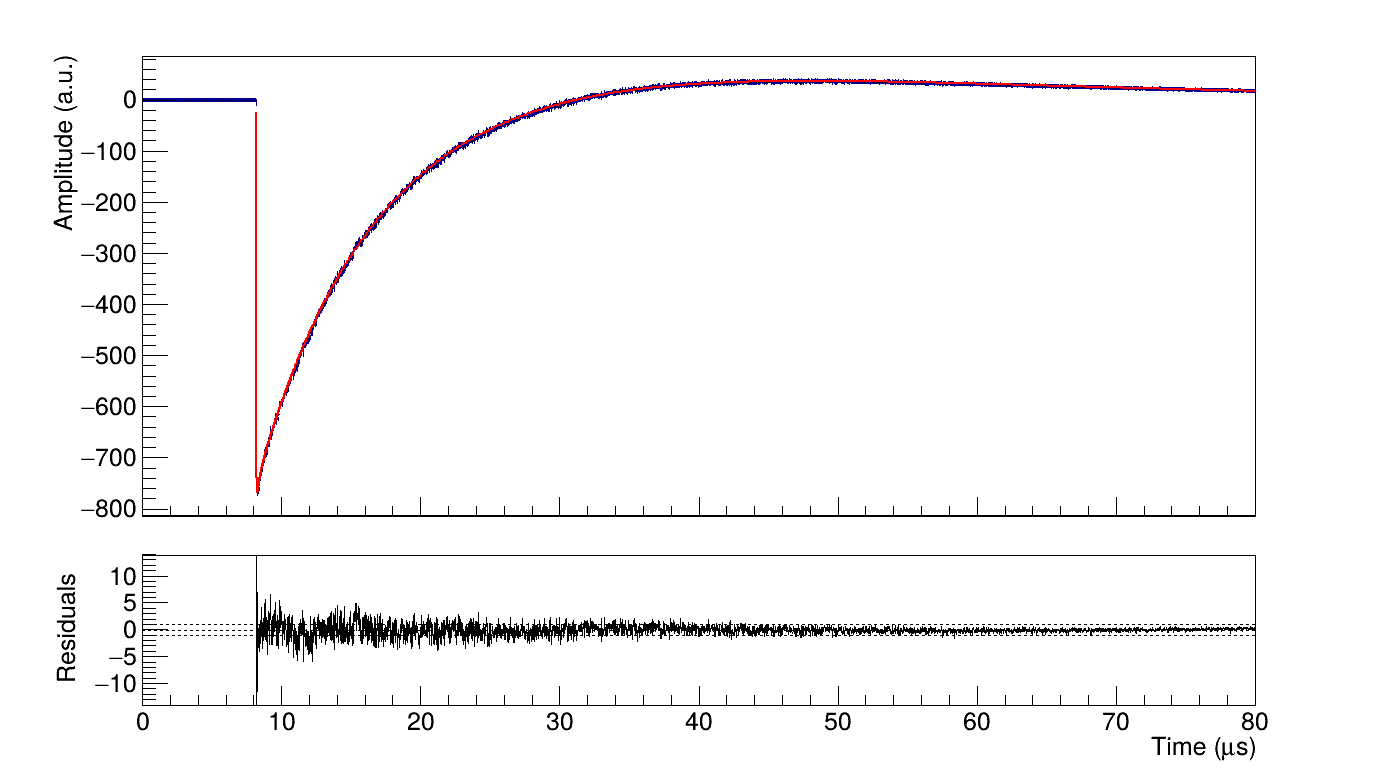} 
 \caption{{\it Top:} Example of scintillation pulse profile of an event with energy 200.6 keV
recorded by the 14-bit digitizer over a time window of 80 $\mu$s. The amplitude is in
arbitrary unit. The photoelectrons of the pulse are resolved by the electronics.
{\it Bottom:} Average pulse profile calculated by considering 4000
scintillation pulses with energy in the range (190--210) keV; in the
plot the red curve is the result of the fit (see text).}
 \label{pulse}
\end{figure}

The PMT waveform (see Fig. \ref{pulse}-top) was digitized using a CAEN DT5724 14-bit digitizer (100 MSamples/s), which 
recorded the scintillation pulses event-by-event within a time window of 80 $\mu$s. 
No significant dead time is present in the measurement; 
in particular, the digitizer allows buffering 50 events and this assures 
a dead-time-free data acquisition. 
Only events occurring within the time window of a previous event are discarded in the analysis, although they are acquired.
This effect introduces a correction of $\sim 1\%$ in the live time of the production runs.
In Fig. \ref{pulse}-bottom
the average pulse profile, calculated by considering 4000
scintillation pulses with energy in the range 190-210 keV, is shown. In this
plot the red curve is the result of the fit performed 
from the beginning of the pulse up to the end of the time window and
considering
the sum of three exponential decay components and accounting also for the small overshoot of the pulse 
($\chi^2$/\text{d.o.f.} = 0.93). This overshoot is due to the voltage divider design optimized for positive high voltage power supply.

The residuals, calculated by subtracting the histogram to the fit function and 
dividing for the histogram bin errors, are also reported. 
A primary exponential component was identified, with
a time decay of 13.6(16) $\mu$s, contributing 95.3(17)\% of the total pulse. A second minor component was also detected, with
decay times of 5.1(3) $\mu$s,
contributing 4.7(17)\% of the total. Moreover, a third shorter component (0.226(10) $\mu$s) is necessary ($\simeq 0.05 \%$), which however can be
ascribed to
possible systematical effects in the fit.

The pulses profiles recorded by the DAQ  during the calibration and production runs have been analysed  in order to select 
a clear sample of scintillation profiles. The aim of this analysis was the rejection of noisy events, pile-ups 
and further determination of the spectrum energy threshold. 
In particular, we rejected events having: i) anomalous baseline value measured before the starting 
point of the scintillation profile in the recorded time window (noisy events); ii) anomalous baseline 
fluctuation (produced by noise); iii) large {\it mean time} (pile-up events).
For each event 
the pulse {\it mean time}  \footnote{The {\it mean time} is not directly connected 
to the scintillation decay times, but it depends on them.} and the pulse {\it area} (proportional to the released energy) were estimated over a time window of 5 $\mu$s 
following a method similar as in Ref. \cite{174Hf}. This time  
window was selected starting from the beginning of the pulse until the overshoot effects (see Fig. \ref{pulse}) become significant.
A data analysis procedure to remove the overshoot part of the signal from each scintillation pulse before
calculating the pulse {\it mean time} and the pulse {\it area} has been also performed.
The result of this procedure is discussed in Section \ref{sec_beta}.

\subsection{Energy calibration}
\label{sec_cal}

\begin{figure}[!ht]
\centering
\includegraphics[scale=0.28]{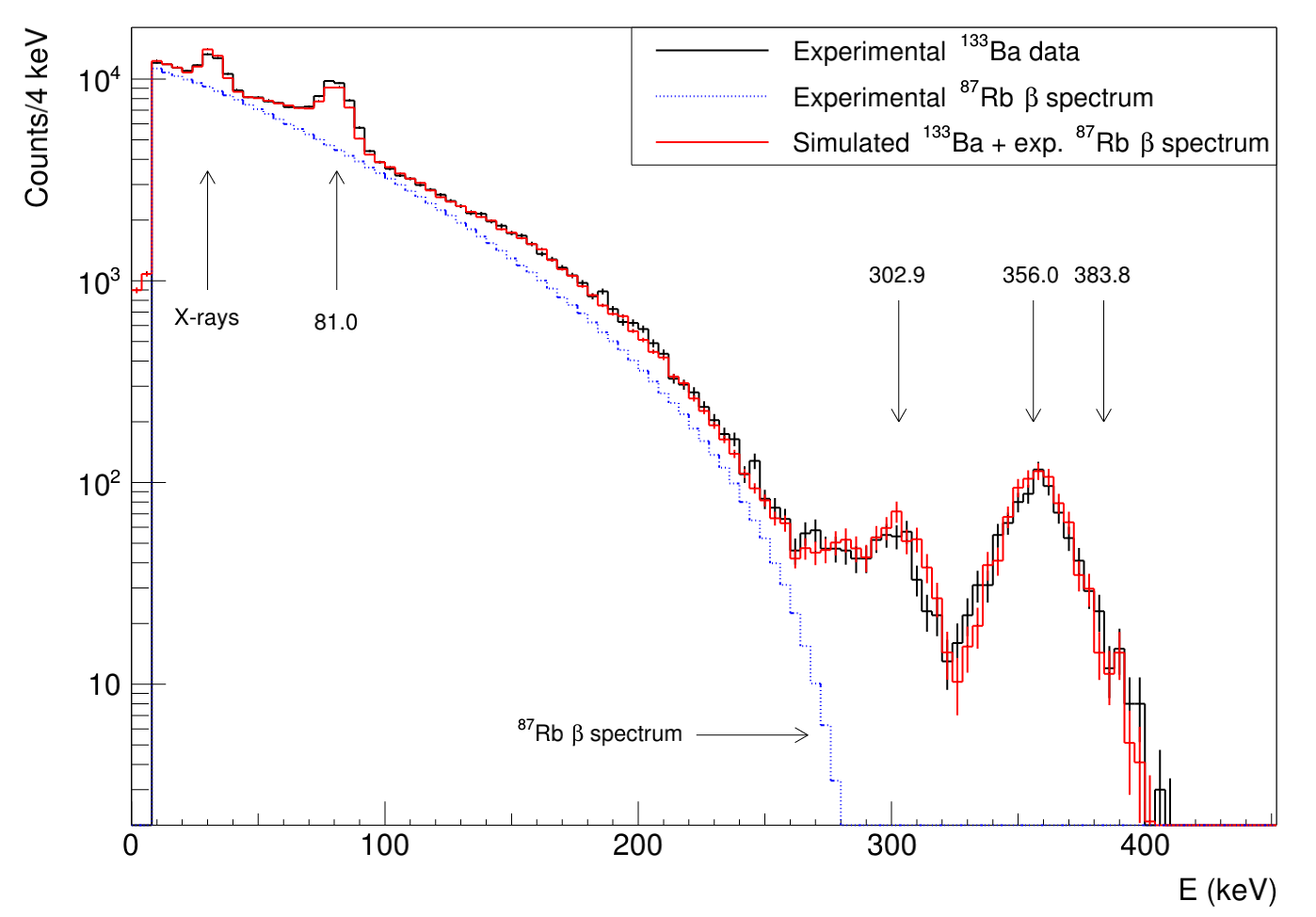} 
 \caption{The experimental spectrum (black histogram) measured in presence of the $^{133}$Ba source 
is shown alongside the whole simulated model (red histogram). This simulation was performed considering a PTFE source holder
with 1 mm thickness and a $^{133}$Ba source wire with 2 mm diameter, 
which provided the best agreement 
with the experimental data. Additionally, the measured $\beta$ spectrum from $^{87}$Rb (see later) is also displayed (blue histogram). The given labels
represent the energy in keV of the main $^{133}$Ba gammas or the X-rays from the daughter  $^{133}$Cs.}
 \label{source}
\end{figure}

Energy calibration and energy resolution of the detector were monitored under the same experimental conditions as the production run by
using a $^{133}$Ba calibration source in form of wire (activity of 65 kBq at the time of measurements), which was positioned close to the detector
via a PTFE tube (see Fig.  \ref{fig_schema}). 
Owing to the large counting rate
of the detector due to its internal
$^{87}$Rb radioactivity, the calibration spectrum obtained irradiating the detector with the $^{133}$Ba calibration source has
a significant contribution from $^{87}$Rb $\beta$-decay, as shown in Fig. \ref{source}.
In fact, the counting rate of the calibration run is 157 Hz ($^{133}$Ba+$^{87}$Rb), 
while it is 123 Hz ($^{87}$Rb) without the source.

Dedicated Monte Carlo simulations were carried out with the Geant4 software package \cite{Geant4} version 4.10.06.p01. 
All the main components of the experimental setup have been implemented in the simulation geometry: the crystal, the PMT, the PTFE supports 
and source holder, the copper shield and the lead shield (sketch and 3D view in Fig. \ref{fig_schema} are obtained from the geometry implemented in Geant4). 
The Livermore physics list was used to describe the electromagnetic interactions. Livermore model relies on an extensive library of evaluated low-energy data and is considered to be one of the most precise and realistic at low energy (recommended validity range: 250 eV - 100 GeV). The range cuts for secondary particles production were reduced to values smaller than the default: 1 nm for electrons/positrons and 1 $\mu$m for gammas. Atomic relaxation processes following ionization were switched on. In particular, fluorescence, Auger effect and particle induced X-Ray emission were enabled and the production threshold was disabled in case of fluorescence and Auger electron production. Finally, the main information collected by the simulation during the particle tracking are the energy depositions in the setup volumes, their delay from the primary event and the nature of the involved particles. 
The Geant4.10.06.p01 radioactive decay and photon evaporation data were used to generate the $^{133}$Ba decays in the calibration source and the decays 
of the radioactive contaminations in the setup. 
The
simulations helped refining the geometric parameters of the setup, such as the size of the source wire and 
the thickness of the PTFE source holder. 
The irregular shape of the used crystal (maximum sizes approximately of $9 \times 6 \times 4$ mm$^3$) was reconstructed; in addition, we have verified that, using a different shape of the crystal 
as a cube of the same mass and edge of 5 mm, the simulated $\beta$-spectra of the two crystal's geometries are rather similar within 0.2\%.

The result of the $^{133}$Ba simulation summed up with the $^{87}$Rb contribution (red histogram online) was
compared with 
the experimentally measured spectrum in presence of the $^{133}$Ba source (black histogram) in Fig. \ref{source}, to determine the energy threshold and the energy scale for the measurement. 
The experimental spectrum was calibrated by a linear function.
Also shown is the measured spectrum without  $^{133}$Ba source, mainly due to the $^{87}$Rb decay as
mentioned previously,
properly normalized (blue histogram online). 
The calibration procedure involved the following steps: i) the simulated $^{133}$Ba 
spectrum was fitted  to the experimental data using the peak at $\simeq$ 360 keV where the contribution of the $^{87}$Rb is absent; 
ii) the experimental $\beta$ spectrum 
(see also later), which served as background in the $^{133}$Ba calibration, was rescaled using the residuals 
from the $^{133}$Ba spectrum in the energy range [100 -- 150] keV. 

Fig. \ref{source} demonstrates good linearity of the RZC
detector's response and shows an energy threshold down to 10 keV. 
Additionally, the energy resolution used in the simulation to match the experimental data
follows the relation: 
FWHM[keV] = 1.41 $\times \sqrt{E[keV]}$ (i.e., FWHM = 4.5 keV or 45\% 
at energy of 10 keV, and FWHM = 25.2 keV or 7.9\% at 320 keV).

\subsection{Data taking and background modelling}
\label{sec_anal}

The measurement of $^{87}$Rb $\beta$-decay was conducted over a live time 
of $t_{live} = 80418.8$ s; i.e. -- owing to the large counting rate of the detector due to the $^{87}$Rb radioactivity -- 
the measurements have been performed in a run over about one day.

\begin{figure}[!h]
\centering
\vspace{-0.4cm}
\includegraphics[width=0.77\linewidth]{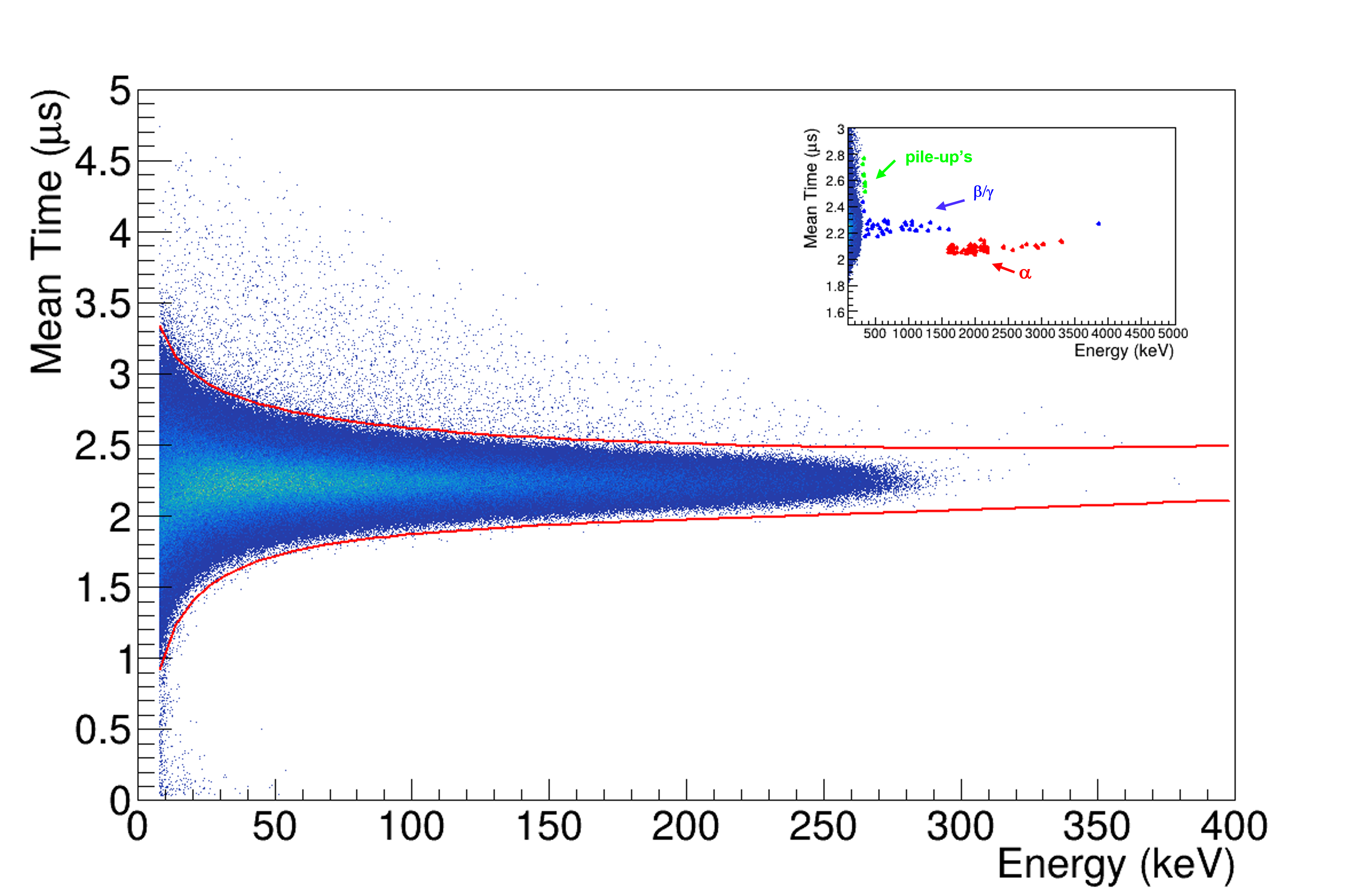}
\caption{Mean time of the pulses measured in a 5 $\mu$s time window by 
the RZC detector for all the collected events as a function of their energy. 
The red curves correspond to the 4$\sigma$ intervals around the  {\it mean time} values 
of scintillation pulses for {\it $\beta/\gamma$ band} events. 
Mainly pile-up events are ruled out by removing events out of the region delimited by the 
red curves. This procedure implies a negligible dead-time of about 0.1\%.}
\label{fig:mean}
\end{figure}

In Fig. \ref{fig:mean} the scatter plot of the pulse {\it mean time} as a function of the energy is reported for all
the events in the data set. The events above and below the red curves were rejected. The
selection lines are the 4$\sigma$ intervals around the  {\it mean time} values of scintillation pulses
for beta/gamma events ({\it $\beta/\gamma$ band}). 
From this selection procedure a cautious energy threshold of 10 keV has been derived. 
The number of rejected events above 10 keV is very low: 11\,577, i.e., 0.1\%, 
and the live time reduction of the data after this selection corresponds to 0.1\%. 
The number of accepted events above 10 keV is about $10^7$. 
Only 37 events in the {\it $\beta/\gamma$ band} are observed above 320 keV,
slightly above
the experimental end point
of the $\beta$-decay of $^{87}$Rb which includes the role of the energy 
resolution of the RZC detectors.
These latter events are neither pile-up nor noisy events. The time of occurrence of these events is uniformly distributed over the total running time.

\begin{figure}
\centering
\includegraphics[width=0.475\linewidth]{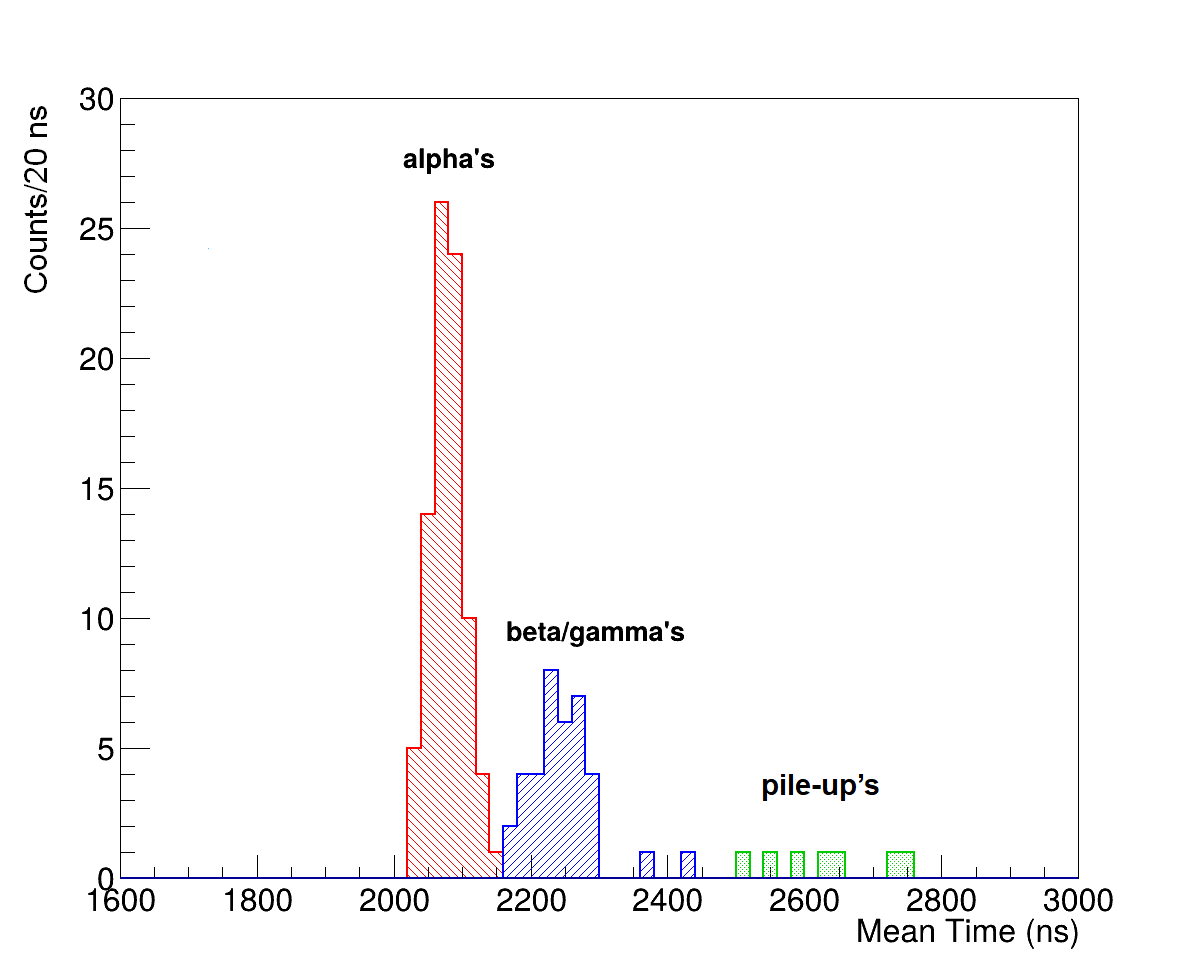}
\includegraphics[width=0.50\linewidth]{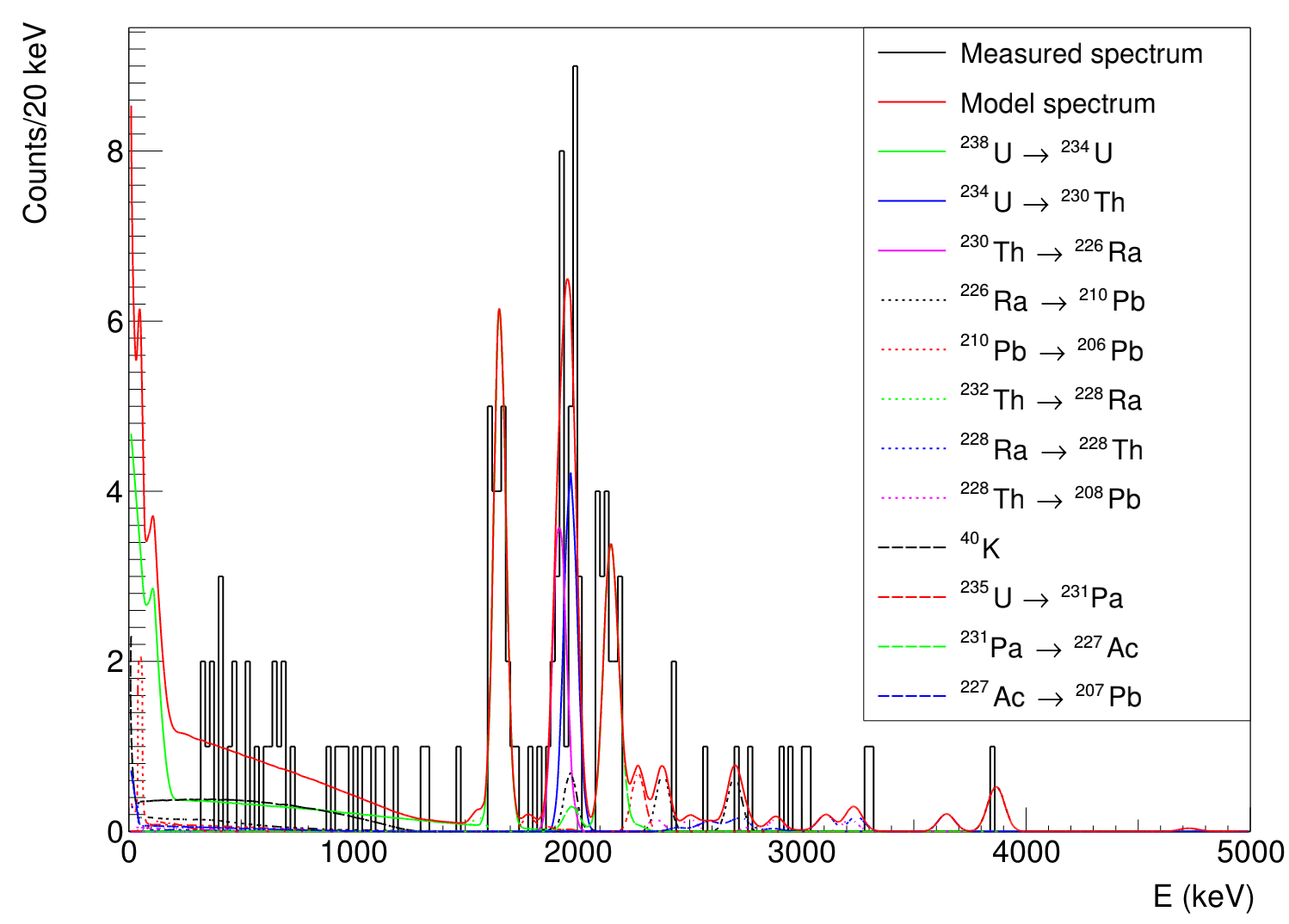}
\caption{{\it Left}:  {\it mean time} distributions of events with energy above 320 keV. Three populations are
shown: the red histogram has been obtained by considering events below the
lower red curve of \mbox{Fig. \ref{fig:mean}} (84 $\alpha$ events);
the blue histogram has been obtained by considering  events in between
the red curves of Fig. \ref{fig:mean} (37 $\beta/\gamma$ events); the green histogram
are events above red band mostly due to pile-ups
(7 events).
{\it Right}: the energy spectrum of the {\it $\beta/\gamma$ band} and $\alpha$ events measured above 320 keV and the background model (red line),
 based on the ICP-MS data reported in Table \ref{conc}, 
 considering in particular the 
contributions from $^{238}$U, $^{232}$Th and $^{235}$U chains and from $^{40}$K.}
\label{fig:psd}
\end{figure}

A population of events around 2 MeV outside the {\it $\beta/\gamma$ band} with shorter  {\it mean time} is observed.
These events are attributed to background $\alpha$ particles.   
In Fig. \ref{fig:psd}-{\it left} the  {\it mean time} distribution of all collected
events above 320 keV is shown. Three populations are
present: i) the red histogram has been obtained by considering events below the
lower red curve of \mbox{Fig. \ref{fig:mean}} (84 $\alpha$ events); ii) 
the blue histogram has been obtained by considering  events in between
the red curves of Fig. \ref{fig:mean} (37 $\beta/\gamma$ events); iii) the green histogram
are events above red band mostly due to pile-ups
(7 events).
The $\alpha$ events correspond to an $\alpha$ activity of
2.25(25) Bq/kg; actually, this value is rather large (in CHC and CZC crystals recently developed it was 100 times 
smaller, even less \cite{Hf1,Hf2,Zr1,Zr2}). This can demonstrate how much the multi-fold vacuum sublimation 
is effective in U/Th
reduction, since for the crystal used here only a single-fold sublimation
of the initial ZrCl$_4$ powder has been performed.

Fig. \ref{fig:psd}-{\it right} shows the energy spectrum of the $\beta/\gamma$ and ${\alpha}$ events measured above 
320 keV and the background model (red line) based on the ICP-MS data of
contamination elements reported in Table \ref{conc}.
The study of the background allows us to obtain the $\alpha/\beta$ ratio for alpha particles in the energy range from 4.0 to 8.0 MeV: 
$\alpha/\beta \approx 0.259 + 0.0315 \cdot E_{\alpha}$[MeV], where the energy of the $\alpha$ particles, $E_{\alpha}$, is expressed in MeV. 
The $\alpha/\beta$ ratio allows the reconstruction of the $\alpha$ energy in keV electron equivalent (keVee).
According to the background model, $\approx 45$ background events are expected at energies below 320 keV 
during the data taking.
Therefore, the background contribution in the energy region of the $^{87}$Rb $\beta$-decay 
is negligible (S/B $\sim 2\times10^5$)
and cannot distort the measured $\beta$ spectrum.

\section{Measurements of $^{87}$Rb $\beta$-decay}
\label{sec_beta}

\begin{figure}[!htp]
\centering
\includegraphics[scale=0.25]{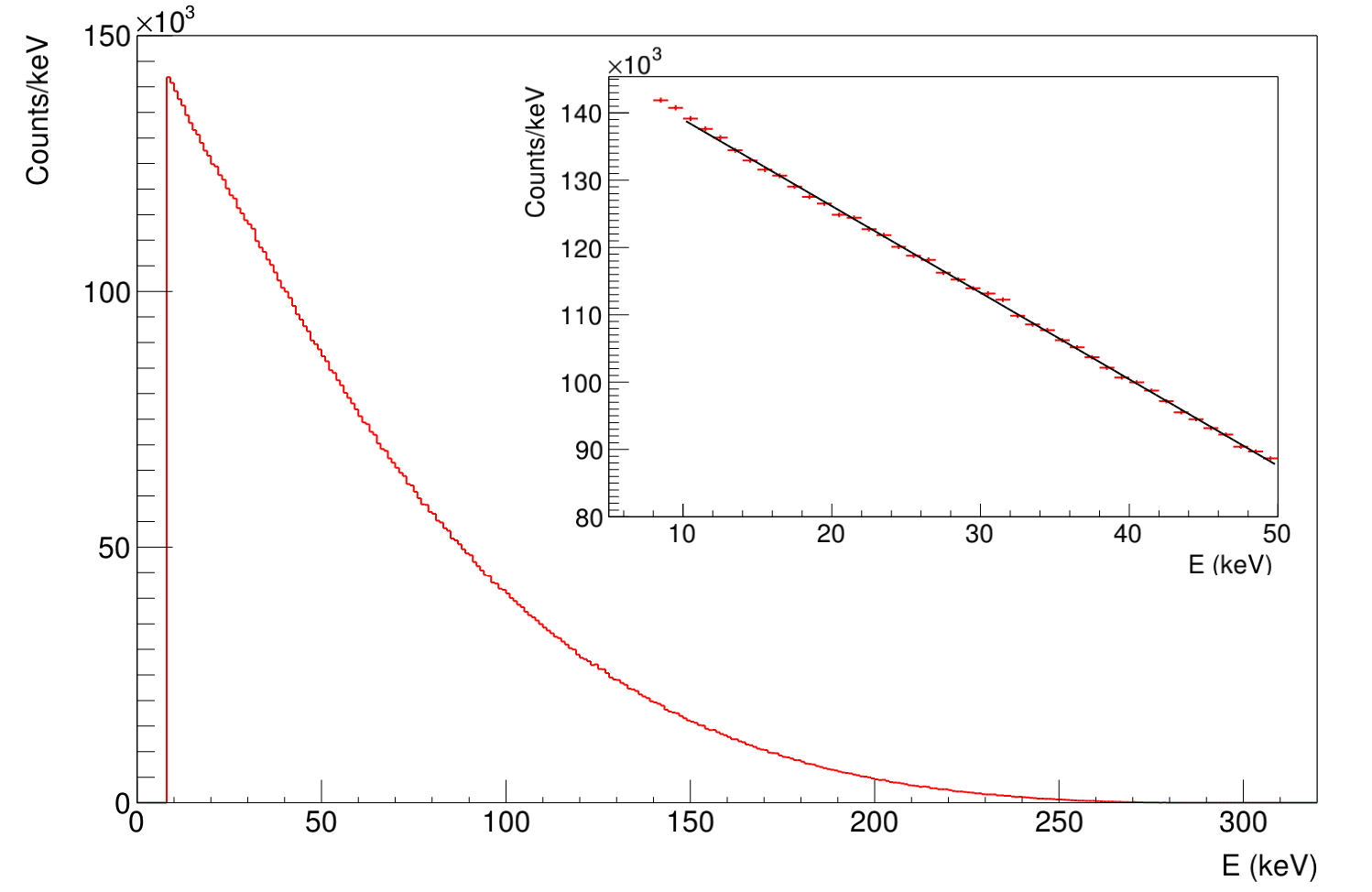} 
\caption{Measured energy spectrum of the events in the  {\it $\beta/\gamma$ band}. In the inset:
the spectrum in the region [10, 50] keV. The latter further supports the 10 keV energy threshold.
In fact, no change of slope is present and a $\chi^2 /\text{d.o.f.}= 42/38 $ (p-value=30\%) is obtained 
when fitting the data (red points online) with a straight line (black online).}
 \label{thr}
\end{figure}

The measured energy spectrum of the events in the  {\it $\beta/\gamma$ band}  is given in Fig. \ref{thr}.
The spectrum in the region [10, 50] keV is given in the inset with superimposed a linear fit
to point out a possible energy threshold effect. Since no change of slope is observed,
the energy threshold of 10 keV can safely be adopted.

The presence of the overshoot in the scintillation pulse profiles
could affect the determination of the total area of each pulse, i.e., the energy of the event.
This effect becomes increasingly significant as it approaches the energy threshold.
For this reason a different approach has been also pursued to evaluate the impact of the overshoot
in the determination of the experimental spectrum.  
In particular, an algorithm was developed to subtract
the overshoot component for each pulse, before calculating the pulse area. 
\begin{figure}[!ht]
\centering
\includegraphics[scale=0.25]{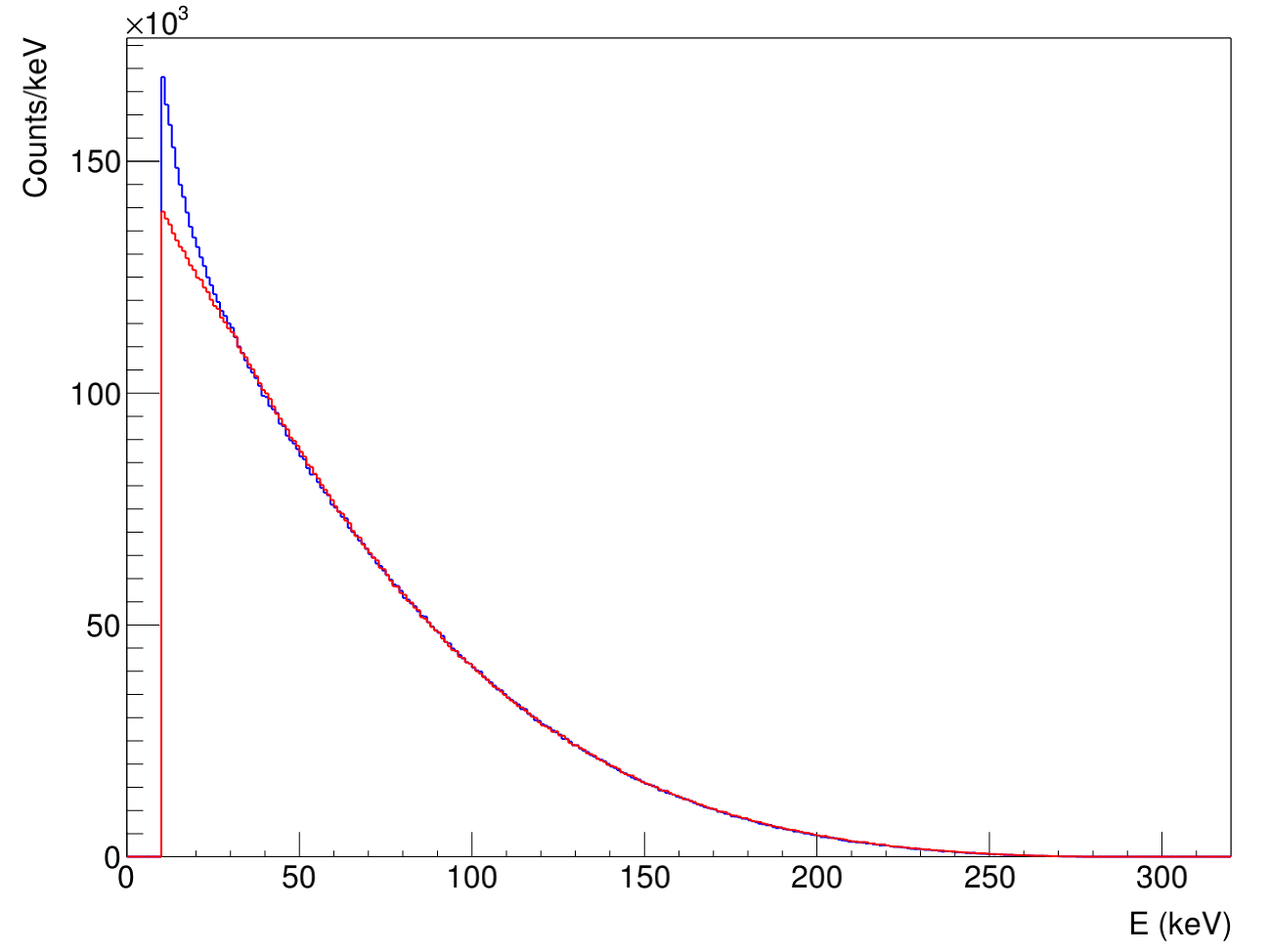} 
 \caption{Experimental energy spectra of the RZC detector obtained by considering two approaches in the data analysis:
 i) scintillation pulses area evaluated in a $5\;\mu$s  time window, no 
 overshoot subtraction algorithm applied and 
 calibration of the energy scale with linear function (red histogram);
 ii)  scintillation pulses area evaluated in a $5\;\mu$s time window, 
 overshoot subtraction algorithm applied and calibration of the energy scale 
 with second-degree polynomial function (blue histogram).}
 \label{blu-red-spect}
\end{figure} 
Using this approach, the energy spectrum does not depend on the chosen width of the
integration time for energy evaluation. Nevertheless, an integration time window of 5 $\mu$s was also used in this case.
In fact, the short time window allows minimizing the role of the overshoot (even if subtracted with some 
procedure). Moreover, 
as mentioned above, the analysis with a shorter gate reduces the number of events rejected due to pile-up, without 
causing significant degradation in energy resolution, also considering that the $\beta$ spectrum is continuous. 
In addition, shorter gate avoids to include in the analysis any ``periodic trains of noise" 
that could affect specially the low energy part.

The two main differences 
with the case of no overshoot subtraction are: i) a 
slightly improved energy resolution and ii) a deviation of the energy scale from the linear trend at 
E$<$100 keV. In particular, the deviation is about 7.5\% at 30 keV where the structure due to the X-ray lines 
from $^{133}$Ba source is present. This deviation was corrected by applying a second-degree polynomial function 
for the calibration of the energy scale: E [keV] = $3.3 + 0.97 \times A + 0.625 \times 10^{-4} \times A^2$, 
where $A$ is the area of the pulse normalized to have $A=356.0$ at deposited energy equal to 356.0 keV (see Fig. \ref{source}). 
This law holds up to $\simeq 500$ keV. 

The energy spectra obtained by considering the two approaches are reported in 
Fig. \ref{blu-red-spect}. The two histograms 
have been obtained in the case of: i) scintillation pulses area 
evaluated in a $5\;\mu$s time window without applying overshoot 
subtraction algorithm and with linear function for the calibration of the energy scale (red histogram);
ii)  scintillation pulses area evaluated in a $5\;\mu$s time window 
applying overshoot subtraction algorithm and with second-degree polynomial function
for the calibration of the energy scale (blue histogram). The two spectra are overlapping for energy 
above $\simeq 30$ keV, while an increase of the blue histogram is evident at lower energy. Thus, 
they can be considered as upper and lower limit, and their spread as the uncertainty associated to the spectrum 
below 30 keV.

\subsection{The detection efficiency for $\beta$-decay from $^{87}$Rb}
\label{sec_eff}

\begin{figure}[!htp]
\centering
\includegraphics[scale=0.24]{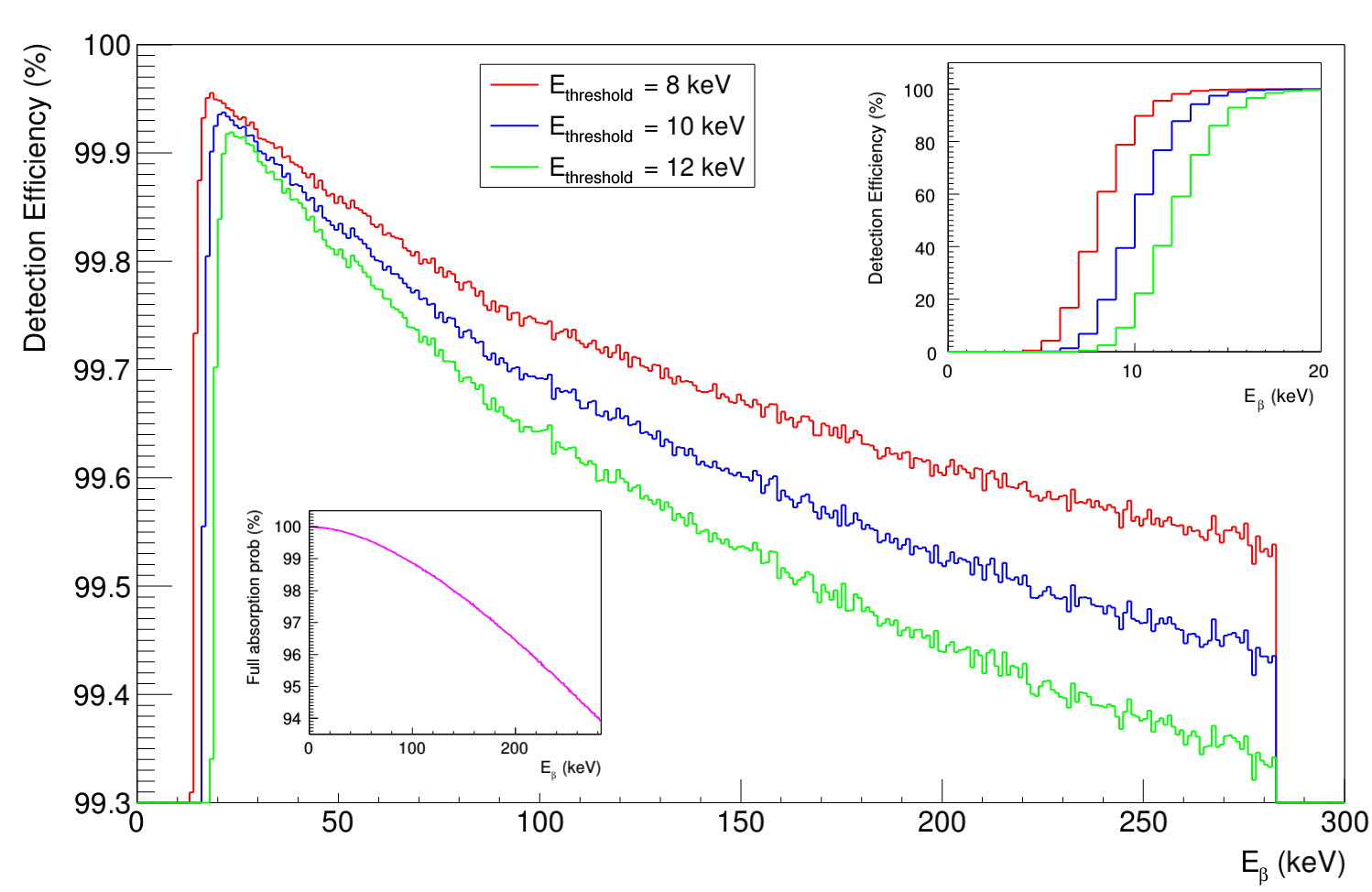} 
 \caption{The detection efficiency as a function of the kinetic energy, E$_{\beta}$, of 
 $\beta$ events in the range of the $^{87}$Rb decay (Q$_{\beta}=282.275(6)$ keV) is examined
 by Monte Carlo simulations
 for three different energy thresholds. 
 The effect of the energy threshold on detection efficiency is particularly evident at low energies, as
 highlighted in the inset on the top. In this work the adopted energy threshold is 10 keV.
 The other inset shows the full-absorption probability calculated by Monte Carlo simulations as a function of E$_{\beta}$.
 }
 \label{eff}
\end{figure}

The detection efficiency for $\beta$-emitters uniformly distributed throughout the detector volume, 
depends on several factors: i) the detector material; ii) the detector shape; iii) the detector size; 
iv) the energy resolution; and v) the energy threshold. 
In this measurement, the detection efficiency was estimated using Monte Carlo simulations, where a $\beta$-decay 
was considered detected if it produced a detected energy -- after broadening with the energy resolution dependence --
 larger than or equal to the energy threshold. 
Fig. \ref{eff} illustrates the calculated detection efficiency of the RZC detector as a function of the kinetic 
energy, $E_{\beta}$, of $\beta$ events in the range of the $^{87}$Rb decay (Q$_{\beta}=282.275(6)$ keV);
in the inset on the top the energy region below 20 keV is pointed out. In particular, in this figure 
the detection efficiency is examined by Monte Carlo simulations for three different energy thresholds. The
effect of the energy threshold on the detection efficiency is particularly evident at low energies, as
highlighted in the inset. For the 10 keV energy threshold adopted in this work the efficiency is $<$ 99.3\% at E$_\beta < $16 keV (see the inset) and  
quickly decreases lowering the energy because of the finite energy
resolution, that shifts some of the events below the adopted energy threshold.
This effect is not visible in the measured data (see Fig. \ref{thr}) because it is compensated by events with energies below the energy threshold,
but whose detected energies exceed it due to the energy resolution.
The other inset of Fig. \ref{eff} shows the full-absorption probability calculated by Monte Carlo simulations as a function of E$_{\beta}$.

\subsection{The unfolding of the $\beta$ spectrum}
\label{sec_unfold}

The real $\beta$ spectrum of the $^{87}$Rb decay was retrieved by applying an unfolding procedure to the
two histograms presented in Fig. \ref{blu-red-spect}, considering them as two extreme cases. 

In detail, the measured $\beta$ spectrum is expected to be different from the 
real $\beta$ spectrum produced by the $^{87}$Rb decay.
This is mainly due to the finite energy resolution of the detector (see Sect. \ref{sec_cal})
and to its detection efficiency $<100$\% (see Sect. \ref{sec_eff}).

The response of the used RZC detector to $\beta$ particles evenly distributed 
in its volume was calculated using Geant4
code and accounting for the finite energy resolution of the detector measured with $\gamma$ sources.

The energy range between 0 keV and the $\beta$ end-point of $^{87}$Rb decay (Q$_{\beta}=282.275(6)$ keV) was 
divided in steps of 1 keV and the response functions were calculated for each of them: $R_i (E_{det})$, $i=1,282$.
In particular, $R_i (E_{det})$ are the calculated energy spectrum of the detected energies, $E_{det}$, in case 
of $\beta$ particles with initial energy E = ``$i$'' keV and starting points uniformly distributed in the detector.
Then, the response functions $R_i (E_{det})$ were normalized dividing by the number of simulated events, $N_i$, 
in each energy step ($\sim 10^6$ simulations/keV): $r_i (E_{det}) = \frac{R_i (E_{det})}{N_i}$.
Finally, the normalized response functions $r_i (E_{det})$ were linearly combined with different weights $s_i$ in 
order to reproduce the measured $\beta$ spectrum: $m(E_{det}) = 
\sum_{i=1}^{282} s_i \times r_i (E_{det})$.
The $s_i$ values were determined with successive approximations\footnote{At each step the ratio between 
the measured and simulated spectra is calculated and fitted with a polynomial function. Then, the fit result is used to rescale 
the $s_i$ values before a new step. The procedure ends when the simulated spectrum agrees with the measured one within 
the statistical uncertainties.}, 
using as starting values $s_i = c_i$, where $c_i$ 
are the number of counts at $i$ keV collected in the measured $\beta$ spectrum.
The final $s_i$ values, that allow a correct reproduction of the measured $\beta$ spectrum $m(E_{det})$, represent 
the unfolded $\beta$ spectrum, i.e., the real $\beta$ spectrum 
produced by the $^{87}$Rb decay during the measuring time. 

\begin{figure}[!h]
\centering
\includegraphics[width=0.45\linewidth]{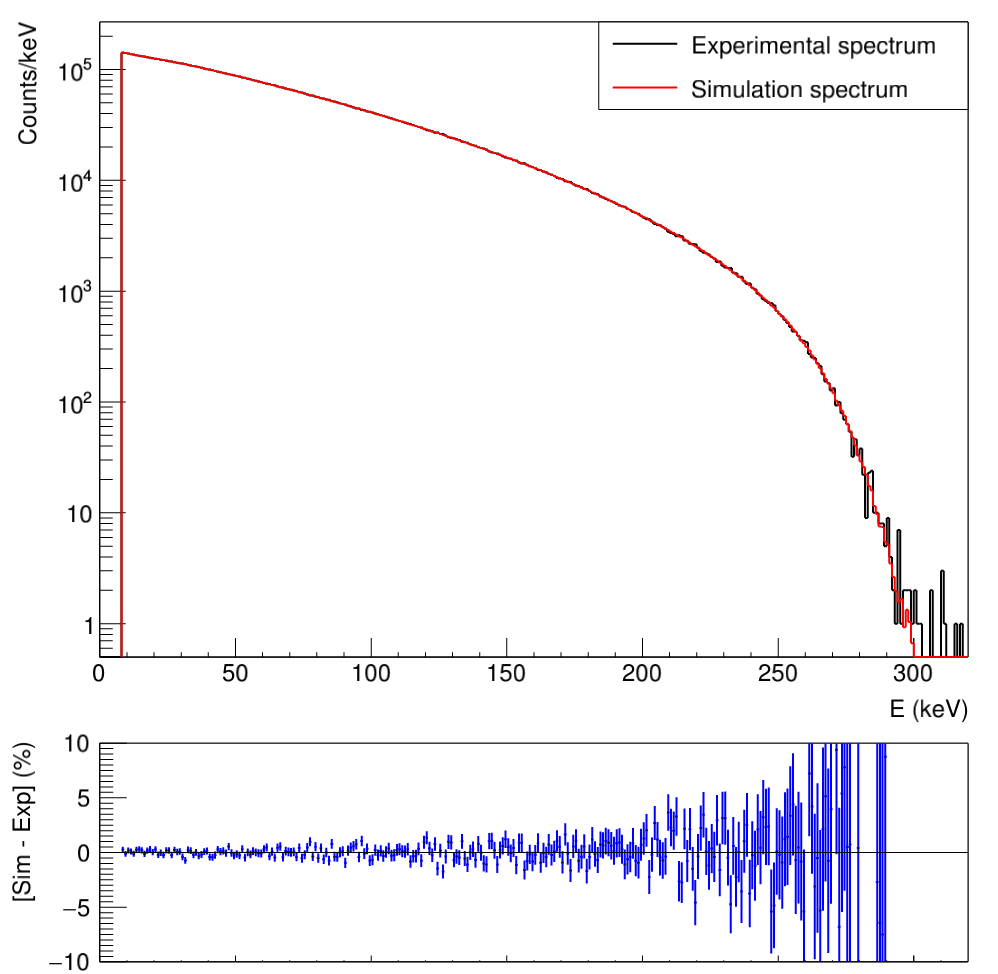}
\includegraphics[width=0.45\linewidth]{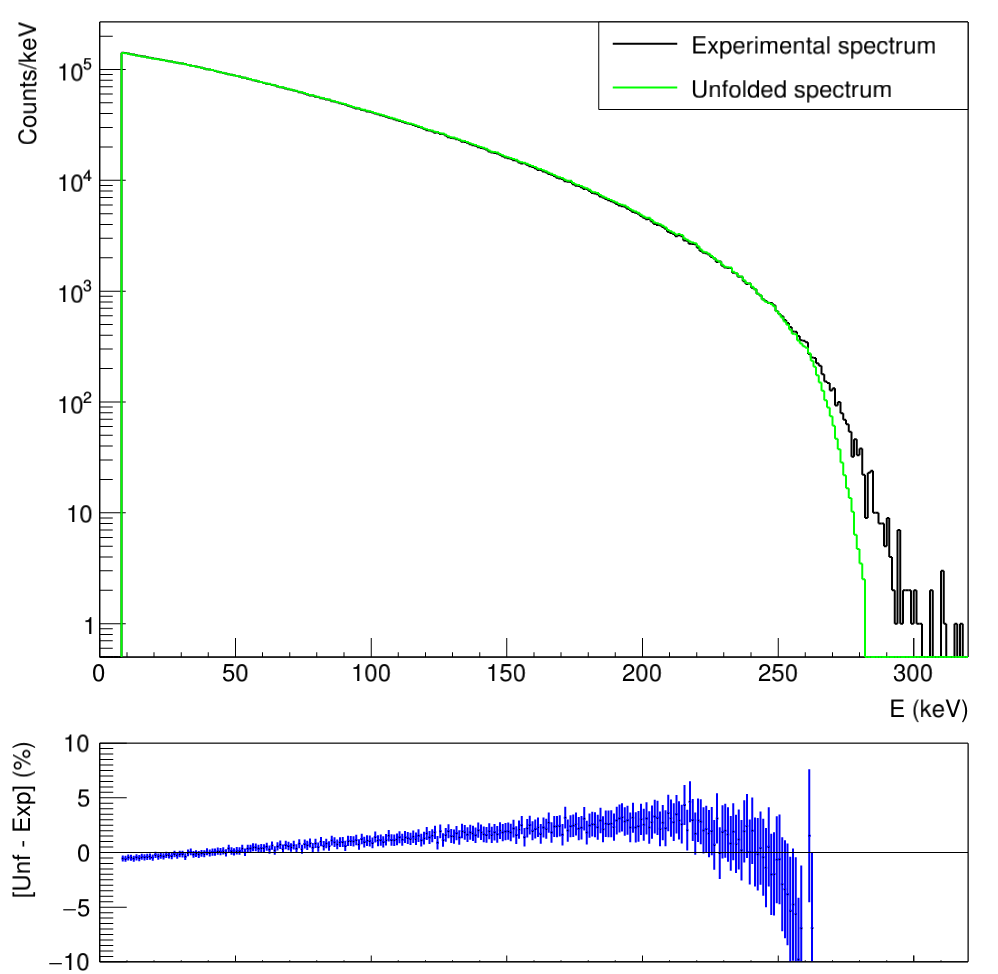}
\caption{{\it Left:} Comparison between the measured $\beta$ spectrum of $^{87}$Rb obtained 
without considering the overshoot subtraction algorithm (red histogram in Fig. \ref{blu-red-spect})
and the simulation spectrum, obtained
using the unfolded spectrum (see text) to describe the starting energy of the $\beta$ events in the RZC detector, and
accounting for the finite energy resolution of the RZC detector.
{\it Right:} Comparison between the measured and the unfolded $\beta$ spectrum of $^{87}$Rb. 
The distortion of the measured spectrum is 
of the order of few \% up to 250 keV; above this energy the distortion due to the energy resolution starts 
to be very important.}
\label{fig-spec}
\end{figure}

As an example, we examine the case of the experimental spectrum obtained 
without considering the overshoot subtraction algorithm (red histogram in Fig. \ref{blu-red-spect}).
As a result, Fig. \ref{fig-spec}-left shows the comparison between the experimental spectrum and the
simulation obtained when the unfolded spectrum is the input of the Monte Carlo simulation and the energy resolution is accounted for.
Thus, the unfolding procedure well reproduces the experimental spectrum both in shape and in counting rate.

Moreover, in Fig. \ref{fig-spec}-right the same experimental spectrum is compared
with the unfolded one: 
the distortion of the measured spectrum is of the order of few \% up to 250 keV.
The distortion cannot be reduced by increasing the statistics because it is an intrinsic effect of the detector, 
due to its detection efficiency and its energy resolution. The latter one is the main responsible of the 
significant distortion at E $>$ 250 keV, as shown in Fig. \ref{fig-spec}-right. 
Similar result is obtained when considering the experimental spectrum from the overshoot subtraction approach.

\begin{figure}[!ht]
\centering
\includegraphics[scale=0.30]{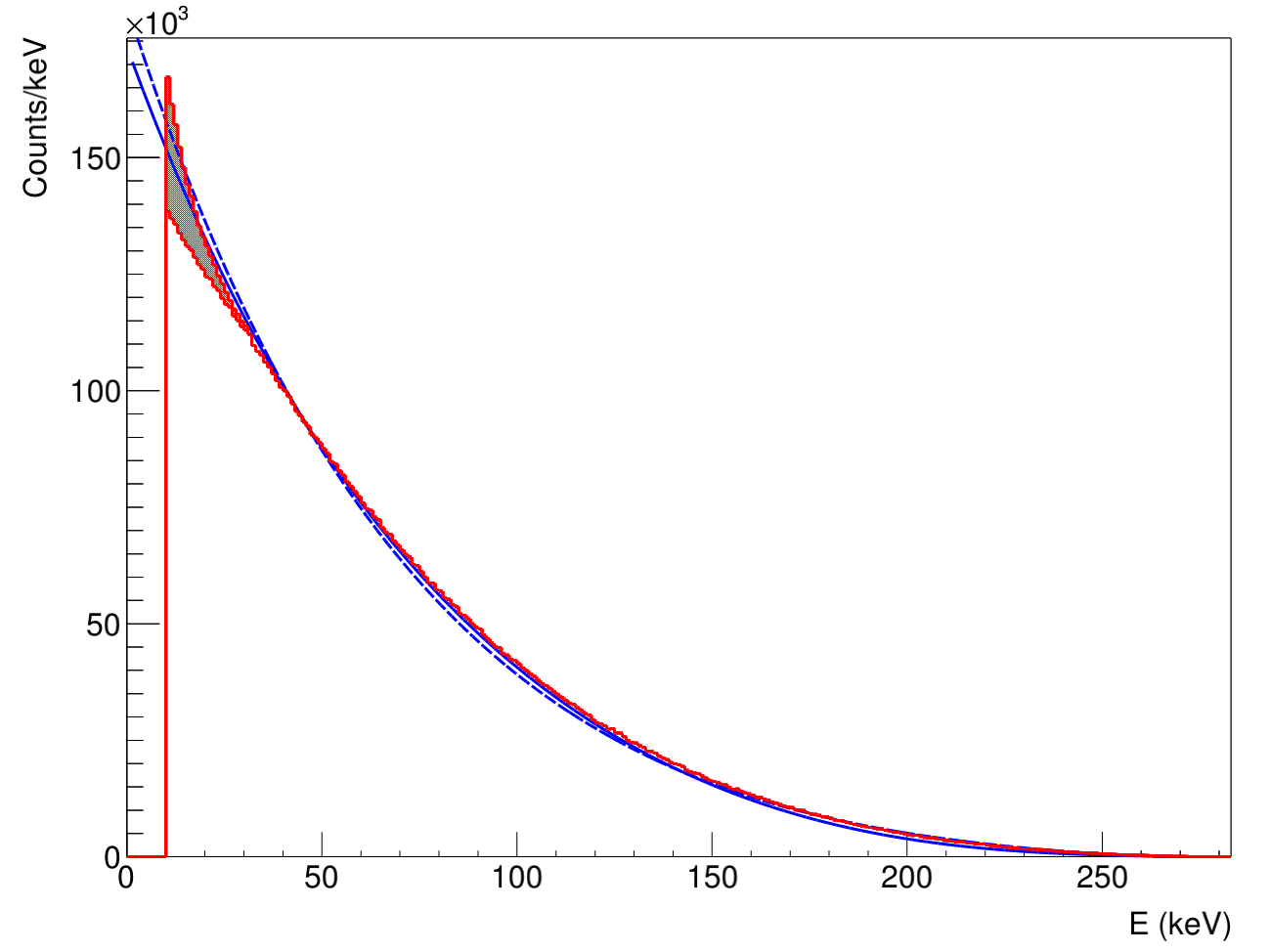}
 \caption{Unfolded energy spectrum for the  $^{87}$Rb $\beta$-decay. The unfolding procedure 
 has been applied to the histograms obtained in the two analysis approaches; the red band accounts for the uncertainty 
 in the original experimental spectrum as a function of the energy. 
 The two blue curves represent: i) the spectrum with the parameters from Ref. \cite{kos2003},
  based on experimental data in the range 6-275 keV,
  and endpoint of 283.3 keV (continuous line); 
 ii) the spectrum with the parameters from Ref. \cite{car2006,car2007},
  based on experimental data in the range 60-180 keV,
  and endpoint of 273 keV (dashed line). See text.}
 \label{spect-band}
\end{figure}

The result of the unfolding procedure is the red band shown in Fig. \ref{spect-band},
accounting for the uncertainty in the original experimental spectrum as a function of the energy.

Comparisons with previous measurements of the $\beta$ spectrum are also shown in Fig. \ref{spect-band}.
To our knowledge, only rather old data are available, the most recent dating back to 2007  \cite{car2006,car2007,rue1973,ege1961,lew1952}. 
A compilation published in 2003 \cite{kos2003}, based primarily on the data of Ref. \cite{rue1973,ege1961}, provided a parametrization of the shape 
factor\footnote{The full parametrization of the $\beta$ spectrum is reported in many textbook;
the Fermi function used here is described e.g. in Ref. \cite{bel07}.}
with an endpoint energy of 283.3 keV:

\begin{equation}
C(W)=q^4 + a p^2q^2 + b p^4 , 
\end{equation}
where $p$ and $q$ denote the electron and neutrino momenta, respectively, and $W$ is the total electron energy, given in units of $m_ec^2$, 
with $m_e$  being the electron rest mass. The fitted values were $a \simeq 0.354$ and $b \simeq 0.003$ \cite{kos2003}.

A subsequent measurement with liquid scintillators in 2007 \cite{car2007} yielded a different endpoint and a new parametrization of the shape factor, with
$a = 0.305(9)$ and $b = 0.011(8)$ \cite{car2006,car2007}.
Both parametrizations are included in Fig. \ref{spect-band}. 
Despite the differences in endpoint energy and fitted parameters, they remain relatively close and are compatible with the present measurement, particularly when considering the age of the data and the distinct experimental systematics involved.

Finally, fitting the present spectrum, with Q$_{\beta}=282.275$ keV \cite{AME2020},
in the energy range 35–283 keV, where the uncertainties are well reduced, yields: $a = 0.3607(10)$ and $b = 0.00620(3)$.

\vspace{5mm}

The number of events of the unfolded spectrum above 10 keV in Fig. \ref{spect-band} is
N$_{decay[10-282]} = 9.742(95) \times 10^{6}$; 
the uncertainty has been estimated considering the red band in Fig. \ref{spect-band}. 
This means that an uncertainty $<1\%$ is obtained 
in the number of decays in the two analysis approaches.
The half-life of $^{87}$Rb $\beta$-decay can be calculated as: 
T$_{1/2} = \ln 2 \cdot  N_{87} \cdot t_{live} $/ N$_{decay}$,
where N$_{decay}$ is given in the two extreme cases by extrapolating the unfolded spectrum down to 0 keV;
using as a case study a straight line:
N$_{decay} = 1.140(29) \times 10^7$.
Therefore, one can obtain: T$_{1/2} = 5.08(13) \times$ 10$^{10}$ yr, 
consistent
with the literature value 
(T$_{1/2} = 4.9650(40) \times 10^{10}$ yr \cite{DDEP2025}).

In the following sections the investigation of the measured unfolded spectrum with theoretical expectations 
will allow us to derive the physical quantities of interest and, in particular,
a new $g_{\rm A}$ determination.

\section{Theoretical background and analyses}

The half-life of a $\beta$ transition can be obtained from 
\begin{equation}
T_{1/2}=\frac{\kappa}{\tilde{C}} \,,
\end{equation}
where $\kappa$ is a constant \cite{Kum2020,Kum2021} and $\tilde{C}$ is the integrated shape function $C$, containing phase-space factors and NMEs, which we calculate to the next-to-leading-order expansion, as discussed in detail in \cite{Haa2016,Haa2017}. Our $\beta$ spectral shape and half-life analyses employ screening, radiative, and atomic-exchange corrections. In the current study, for electron energies below some tens of keV, the $\beta$ spectral shape is mostly dominated by the atomic exchange correction, recently developed 
for allowed $\beta$-decays in \cite{Nitescu2023} and herein used as a surrogate for the third forbidden non-unique decay of $^{87}$Rb. This correction is responsible for the sharp upward trend of the spectral shape at low electron energies.
For the radiative corrections we use the results of the work \cite{Sirl1967}.

The complexity of the shape function $C$ can, however, be condensed to a simple dependence on the weak couplings by writing
\begin{equation}
C(w_e) = g_{\rm V}^2C_{\rm V}(w_e) + g_{\rm A}^2C_{\rm A}(w_e) + g_{\rm V}g_{\rm A}C_{\rm VA}(w_e) \,
\end{equation}
where $w_e$ is the total (rest-mass plus kinetic) energy of the emitted electron. The variation of the shape of the $\beta$-electron spectrum with the value of $g_{\rm A}$ comes from the subtle interference of the combined vector $C_{\rm V}(w_e)$ and axial $C_{\rm A}(w_e)$ parts with the mixed vector-axial part $C_{\rm VA}(w_e)$ \cite{Haa2016}. In some cases (\cite{Eji2019,Ram2024,Kum2020,Kum2021}) this causes measurable changes in the $\beta$ spectra. Thus, the effective value $g_{\rm A}^{\rm eff}$ can be found by computing theoretical $\beta$ spectra for a set of $g_{\rm A}^{\rm eff}$ values and comparing them with the measured one. This method, the spectrum-shape method (SSM), was introduced in \cite{Haa2016}.

An additional contributor to the theoretical analyses of $\beta$ spectral shapes and half-lives is the so-called small relativistic vector NME, sNME \cite{Beh1982}. The sNME gathers contributions outside the valence major shell that contains the proton and neutron Fermi surfaces, which makes its calculation particularly hard for nuclear-theory framework used, e.g., in the present work. Despite its smallness, the sNME can influence the $\beta$ spectral shapes and half-lives quite strongly, see \cite{Kos2021,Kos2023,Ram2024,Kum2021}. 
A ballpark value of the sNME is its CVC value \cite{Beh1982}, recently used for spectral-shape calculations in \cite{Kum2020,Kir2019}. It should be noted, though, that the CVC value represents an  ``ideal'', assuming a perfect many-body theory, which is not the case here. To improve the situation, the experimental works \cite{Kos2021,Ban2024,Pag2024} and theoretical works \cite{Kos2023,Kos2024} used sNME as a fitting parameter, together with $g_{\rm A}^{\rm eff}$. In this way it is possible to reproduce both the measured $\beta$ spectral shape and the experimental (partial) half-life of the $\beta$ transition in  question. This phenomenological
approach was coined Enhanced SSM in \cite{Kum2021} and we adopt it as ESSM in the present work. In fact two solutions of sNME can usually be found for each value of $g_{\rm A}^{\rm eff}$, one closer to [sNME(c)] and one farther from [sNME(f)] the CVC value, as noted in \cite{Kos2023,Kos2024,Ram2024,Ram2024b,Ram2024c}.

The nuclear shell model (NSM) calculations presented in this work were performed using the {\sc kshell} software package~\cite{Shi2019}, employing the following effective interactions: \texttt{glbepn}~\cite{Mach1990}, \texttt{jj44b}~\cite{Mukhopadhyay2017}, and \texttt{jun45}~\cite{Honma2009}. While several other interactions (namely, \texttt{ca48mh1}, \texttt{fpg9ta}, \texttt{glepn}, \texttt{GWBXG}, and \texttt{jj44c}) were also investigated, these alternatives often yielded redundant results or produced theoretical observables (such as level schemes and M1/E2 transition properties) that did not provide a satisfactory alignment with experimental data. Consequently, this study focuses on a curated selection of interactions that best describe the observables detailed in Tables~\ref{tab:levels_table} and \ref{tab:M1_E2_table}. Subsequent discussions will exclusively feature this selected set.

The model space for the \texttt{glbepn} interaction is defined relative to a $^{56}\text{Ni}$ closed core. The valence space for this interaction incorporates the proton orbitals $\pi(1p_{3/2})$, $\pi(0f_{5/2})$, $\pi(1p_{1/2})$, and $\pi(0g_{9/2})$, complemented by orbitals with specified occupation limits: $\pi(2s_{1/2})[0]$, $\pi(1d_{5/2})[0]$, and $\pi(1d_{3/2})[0]$. For neutrons, an analogous set of valence orbitals is employed~$\nu(1p_{3/2})$, $\nu(0f_{5/2})$, $\nu(1p_{1/2})$, and $\nu(0g_{9/2})$ with specific truncations applied to these orbitals as follows: $\nu(2s_{1/2})[2]$, $\nu(1d_{5/2})[2]$, and $\nu(1d_{3/2})[0]$. In this notation, `[N]' signifies a computational truncation, restricting the maximum number of nucleons (N) permitted in that particular orbital. For example, $\nu(1d_{5/2})[2]$ allows a maximum of two neutrons in the $\nu(1d_{5/2})$ subshell (which would normally accommodate up to six), whereas $\pi(1d_{3/2})[0]$ effectively treats the $\pi(1d_{3/2})$ orbital as closed for protons in these calculations. Such truncations are necessitated by the computational challenge of managing the extremely 
large M-scheme dimensions encountered; however, these truncations are justified by the magic number $Z=50$ and $N=50$, although for neutrons we allowed some freedom as described above. Illustratively, for $^{87}\text{Sr}$ calculations using the \texttt{glbepn} interaction with these truncations, the M-scheme dimension reaches approximately $10^{10.1}$.

The \texttt{jun45} and \texttt{jj44b} interactions utilize an identical and untruncated valence model space. For calculations with these interactions, a $^{56}\text{Ni}$ closed core is employed. The valence space above this core comprises the $1p_{3/2}, 0f_{5/2}, 1p_{1/2},$ and $0g_{9/2}$ single-particle orbitals, which are identically available for both protons ($\pi$) and neutrons ($\nu$). No truncations are imposed on the occupation of these valence orbitals, and this model space configuration leads to M-scheme dimensions of approximately $10^{5.7}$, a stark difference that shows the exponential behavior by implementing extra orbitals.

This shared model space forms a common basis with the \texttt{glbepn} interaction, which also includes these four core valence orbitals ($1p_{3/2}, 0f_{5/2}, 1p_{1/2}, 0g_{9/2}$). However, the \texttt{glbepn} model space is extended further by the inclusion of the $2s_{1/2}, 1d_{5/2},$ and $1d_{3/2}$ orbitals. A key distinction, particularly for neutron configurations beyond the $N=50$ magic number, is that the \texttt{glbepn} framework permits the occupation of the $\nu(2s_{1/2})[2]$ and $\nu(1d_{5/2})[2]$ orbitals by up to two neutrons each. These configurations are not accessible within the more restricted valence space of the \texttt{jun45} and \texttt{jj44b} interactions.

\subsection{Observables Comparison}

The predictive power of the employed effective interactions is typically evaluated by comparing calculated excitation energies and spin-parity ($J^\pi$) assignments with experimental data. This comparison for the selected interactions is presented in Table~\ref{tab:levels_table} for both $^{87}\text{Rb}$ and $^{87}\text{Sr}$.

\begin{table}[htbp]
\centering
\footnotesize
\setlength{\tabcolsep}{3pt}
\caption{Comparison of evaluation excitation energies for the first five levels in $^{87}\text{Rb}$ and $^{87}\text{Sr}$ with theoretical predictions from the nuclear shell model interactions (\texttt{glbepn}, \texttt{jj44b}, and \texttt{jun45}), presented in MeV. Experimental data from \cite{IAEA:NDS} is shown without the excitation energy uncertainties (typically a few eV), which were omitted for clarity. Asterisks denote an incorrect ordering of an energy level relative to its adjacent state.}
\label{tab:levels_table}
\begin{tabular*}{\textwidth}{@{\extracolsep{\fill}}lcccc@{\hspace{0.25em}}lcccc@{}}
\toprule
\textbf{$J^\pi_n$} & \textbf{Exp.} & \textbf{\texttt{glbepn}} & \textbf{\texttt{jj44b}} & \textbf{\texttt{jun45}} &
\textbf{$J^\pi_n$} & \textbf{Exp.} & \textbf{\texttt{glbepn}} & \textbf{\texttt{jj44b}} & \textbf{\texttt{jun45}} \\
\cmidrule(r{0.125em}){1-5} \cmidrule(l{0.125em}){6-10}
\multicolumn{5}{@{}c@{\hspace{0.25em}}}{\textbf{$^{87}$Rb Levels}} & \multicolumn{5}{c@{}}{\textbf{$^{87}$Sr Levels}} \\
\midrule
$3/2^{-}_{1}$ & $0.000$ & $0.000$ & $0.000$ & $0.000$ & $9/2^{+}_{1}$ & $0.000$ & $0.000$ & $0.000$ & $0.000$ \\
$5/2^{-}_{1}$ & $0.403$ & $0.620$ & $0.384$ & $0.766^{*}$ & $1/2^{-}_{1}$ & $0.389$ & $0.624^{*}$ & $0.103$ & $0.216$ \\
$1/2^{-}_{1}$ & $0.845$ & $0.823$ & $0.864$ & $0.755^{*}$ & $3/2^{-}_{1}$ & $0.873$ & $0.328^{*}$ & $0.530$ & $0.780$ \\
$3/2^{-}_{2}$ & $1.390$ & $1.682$ & $1.572^{*}$ & $1.640$ & $5/2^{+}_{1}$ & $1.228$ & $1.397^{**}$ & $1.142^{*}$ & $1.163$ \\
$1/2^{-}_{2}$ & $1.463$ & $1.774$ & $1.565^{*}$ & $1.991$ & $5/2^{-}_{1}$ & $1.254$ & $1.244^{**}$ & $0.692^{*}$ & $1.199$ \\
\bottomrule
\end{tabular*}
\end{table}

Overall, the chosen interactions demonstrate reasonable agreement with experimental observations, considering the inherent limitations of the shell model framework. The ground state spin-parities of both the parent ($^{87}\text{Rb}$) and daughter ($^{87}\text{Sr}$) nuclei are correctly reproduced by all models.

More specifically, the \texttt{jj44b} interaction best reproduces the lower-lying states of $^{87}\text{Rb}$. However, it exhibits poorer performance for $^{87}\text{Sr}$, systematically underpredicting the excitation energies of its levels.
In contrast, the \texttt{jun45} interaction shows a notable variance in its performance. While it provides the most accurate overall description for $^{87}\text{Sr}$, it struggles to reproduce the energy levels of $^{87}\text{Rb}$.
Lastly, the \texttt{glbepn} interaction provides a fair level of agreement, though it notably predicts an incorrect ordering for the $1/2^-_1$ and $3/2^-_1$ states in $^{87}\text{Sr}$, it also displays a general tendency to overestimate excitation energies.

A further assessment of the employed effective interactions involves the comparison of predicted static electromagnetic moments, i.e., magnetic dipole ($\mu$) and electric quadrupole ($Q$) moments with available experimental data for selected low-lying states in $^{87}\text{Rb}$ and $^{87}\text{Sr}$ \cite{IAEA:NDS}. Electric quadrupole moments are particularly interesting as they provide direct insights into nuclear deformation, a property that can be challenging to capture within the standard spherical shell model framework fully. While some collective effects influencing deformation might be implicitly incorporated into the effective interaction (e.g., through fitting procedures that reproduce regional observables), the explicit reproduction of deformation-sensitive observables like the electric quadrupole remains a critical test. Consequently, comparing model predictions for these moments with experimental data serves as a test of the interaction's performance and the reliability of the interactions for describing such structural details. This analysis is presented in Table~\ref{tab:M1_E2_table}.

\begin{table}[htbp]
\centering
\footnotesize
\setlength{\tabcolsep}{2.5pt}
\caption{Evaluation of nuclear shell model interactions versus experimental \cite{IAEA:NDS} magnetic dipole moments ($\mu$ in $\mu_N$) and electric quadrupole moments ($Q$ in $e\text{b}$) is presented for specific energy levels in $^{87}\text{Rb}$ and $^{87}\text{Sr}$. The model predictions for electric quadrupole moments incorporate effective charges ($e_{\text{eff}}^{p} = 1.5e$, $e_{\text{eff}}^{n} = 0.5e$), while magnetic dipole moment calculations employed bare nucleon $g$-factors ($g_{l}^{(p)}=1$, $g_{l}^{(n)}=0$, $g_{s}^{(p)}=5.585$, $g_{s}^{(n)}=-3.826$).}
\label{tab:M1_E2_table}
\begin{tabular*}{\textwidth}{@{\extracolsep{\fill}}cccccccccc@{}}
\toprule
\textbf{$J^\pi_n$} & \textbf{$E_x$ (MeV)} & \multicolumn{2}{c}{\textbf{Exp.}} & \multicolumn{2}{c}{\textbf{\texttt{glbepn}}} & \multicolumn{2}{c}{\textbf{\texttt{jj44b}}} & \multicolumn{2}{c}{\textbf{\texttt{jun45}}} \\
\cmidrule(lr){3-4} \cmidrule(lr){5-6} \cmidrule(lr){7-8} \cmidrule(lr){9-10}
& & \textbf{$\mu$} & \textbf{$Q$} & \textbf{$\mu$} & \textbf{$Q$} & \textbf{$\mu$} & \textbf{$Q$} & \textbf{$\mu$} & \textbf{$Q$} \\
\midrule
$^{87}$Rb: $3/2^{-}_{1}$ & $0.000$ & $2.75129(8)$ & $0.1335(5)$ & $3.092$ & $0.144$ & $3.381$ & $0.086$ & $2.412$ & $0.159$ \\
\midrule
$^{87}$Sr: $9/2^{+}_{1}$ & $0.000$ & $-1.09316(11)$ & $0.305(2)$ & $-1.717$ & $0.136$ & $-1.399$ & $0.301$ & $-0.966$ & $0.336$ \\
\phantom{$^{87}$Sr:  }$1/2^{-}_{1}$ & $0.389$ & $0.624(4)$ & \text{--} & $0.457$ & \text{--} & $0.674$ & \text{--} & $0.532$ & \text{--} \\
\bottomrule
\end{tabular*}
\end{table}

A summary of the performance of each interaction concerning these M1 and E2 properties indicates varying degrees of success. The \texttt{glbepn} interaction generally struggles to reproduce the magnetic properties, typically overpredicting the magnetic dipole moments. In contrast, 
it accurately reproduces the $Q$ value for $^{87}\text{Rb}$ but performs poorly for $^{87}\text{Sr}$, which may indicate limitations in its ability to consistently describe deformation-related properties across different nuclei. Regarding the \texttt{jj44b} interaction, it provides a reasonable, albeit approximate, description of the electromagnetic moments in $^{87}\text{Rb}$. For $^{87}\text{Sr}$, it shows good agreement for the electric quadrupole moment of the ground state but tends to substantially overpredict its magnetic dipole moments. Finally, the \texttt{jun45} interaction is most effective at predicting the magnetic dipole moments for both nuclei, closely reproducing most of the observed $\mu$ values for the states considered.

In conclusion, guiding the subsequent analysis and interpretation within this work, it is important to note that despite the \texttt{glbepn} interaction employing a larger model space, its reproduction of the investigated observables was generally less accurate compared to \texttt{jj44b}, and particularly so when compared to \texttt{jun45}. However, no single model is universally superior. The \texttt{jun45} interaction is the most effective at predicting the magnetic dipole moments for both nuclei, while the electric quadrupole moments are best described by \texttt{glbepn} for $^{87}\text{Rb}$ and \texttt{jj44b} for $^{87}\text{Sr}$.

Furthermore, it is pertinent to acknowledge that, regarding the calculation of electromagnetic properties, the specific choice of effective charges and g-factors can influence the agreement with experimental data and, consequently, the identification of a `best-overall' interaction. Within this work, standard {\sc kshell} values for these parameters were used; variations in these choices could potentially alter the relative performance of the models, a factor to consider in broader evaluations.

\subsection{Theoretical predictions and comparisons}

For the selected shell-model interactions, one-body transition densities (OBTDs) were computed for the ground-state to ground-state $\beta^-$ decay of $^{87}\text{Rb}(3/2^-) \rightarrow ^{87}\text{Sr}(9/2^+)$. These calculations were complemented by predictions from the Microscopic Quasiparticle-Phonon Model (\texttt{MQPM}) \cite{Toi1998} and recently applied for the same decay~\cite{Kostensalo2017} to provide a comprehensive assessment of this $\beta$-decay process. This third-forbidden non-unique decay involves dominating rank-3 and rank-4 tensor operators, and their individual contributions to the OBTDs are itemized in Table~\ref{tab:OBD_table}, denoted therein as R3 and R4, respectively. The table presents the major contributing transitions; minor contributions have been omitted for brevity.

Analysis of Table~\ref{tab:OBD_table} reveals that the primary contribution to the OBTDs originates from the $\nu 0g_{9/2} \rightarrow \pi 1p_{3/2}$ transition. This dominance is consistently reproduced by all models investigated. The \texttt{MQPM} calculations include contributions from additional orbitals not present in the shell-model valence space; however, these contributions are found to be minor compared to the dominant transitions listed. This observation suggests that the chosen closed core and valence space configurations for the nuclear shell model calculations are largely sufficient for describing the primary aspects of this decay. These OBTDs form the basis for calculating the nuclear matrix elements (NMEs) governing the $\beta$-decay process, which are subsequently used to determine theoretical $\beta$-spectral shapes and the decay half-life.

\begin{table}[htbp]
\centering
\footnotesize
\caption{Calculated one-body transition densities (OBTDs) for the $\beta^-$ decay from the ground state of $^{87}\text{Rb}(3/2^{-})$ to the ground state of $^{87}\text{Sr}(9/2^{+})$. The contributions from distinct neutron-to-proton orbital transitions are itemized in the table, disaggregated into their rank~3~(R3) and rank~4~(R4) tensor components. Not all inclusive, most minor contributions are omitted. }
\label{tab:OBD_table}
\setlength{\tabcolsep}{3pt}
\begin{tabular*}{\textwidth}{@{\extracolsep{\fill}}lcccccccc@{}}
\toprule
\textbf{Transition} & \multicolumn{2}{c}{\textbf{\texttt{MQPM}}} & \multicolumn{2}{c}{\textbf{\texttt{glbepn}}} & \multicolumn{2}{c}{\textbf{\texttt{jj44b}}} & \multicolumn{2}{c}{\textbf{\texttt{jun45}}} \\
\cmidrule(lr){2-3} \cmidrule(lr){4-5} \cmidrule(lr){6-7} \cmidrule(lr){8-9}
($nlj \rightarrow n'l'j'$) & \textbf{R3} & \textbf{R4} & \textbf{R3} & \textbf{R4} & \textbf{R3} & \textbf{R4} & \textbf{R3} & \textbf{R4} \\
\midrule
$0g_{9/2} \rightarrow 1p_{3/2}$ & $-0.837$ & $0.798$ & $-0.888$ & $0.884$ & $-0.759$ & $0.779$ & $-0.805$ & $0.793$ \\
$0g_{9/2} \rightarrow 1p_{1/2}$ & \text{--} & $0.010$ & \text{--} & $-0.062$ & \text{--} & $-0.175$ & \text{--} & $-0.102$ \\
$1p_{3/2} \rightarrow 0g_{9/2}$ & $0.000$ & $-0.003$ & $0.070$ & $0.037$ & $-0.007$ & $0.032$ & $-0.005$ & $0.034$ \\
$1p_{1/2} \rightarrow 0g_{9/2}$ & \text{--} & $-0.000$ & \text{--} & $-0.024$ & \text{--} & $-0.020$ & \text{--} & $-0.048$ \\
$0g_{9/2} \rightarrow 0f_{5/2}$ & $0.012$ & $-0.002$ & $-0.000$ & $0.001$ & $0.032$ & $-0.043$ & $-0.006$ & $-0.006$ \\
$0f_{5/2} \rightarrow 0g_{9/2}$ & $-0.000$ & $0.002$ & $0.029$ & $-0.012$ & $0.041$ & $-0.025$ & $0.041$ & $-0.014$ \\
$1d_{5/2} \rightarrow 1p_{1/2}$ & $0.010$ & \text{--} & $0.000$ & \text{--} & $^*$ & $^*$ & $^*$ & $^*$ \\
$1p_{3/2} \rightarrow 1d_{3/2}$ & $-0.008$ & \text{--} & $0.000$ & \text{--} & $^*$ & $^*$ & $^*$ & $^*$ \\
\bottomrule
\end{tabular*}
\par\vspace{0.5ex}
\parbox{\textwidth}{\raggedright\textit{\footnotesize $^*$Transition involves orbitals outside the defined model space for this interaction.}}
\end{table}

Following the computation of the NMEs for each model, the half-life was calculated using two approaches for the sNME ($^V\mathcal{M}_{KK-11}^{(0)}$). The first approach uses a value derived from the Conserved Vector Current (CVC) relationship to the large-NME ($\ell$-NME or $^V\mathcal{M}_{KK0}^{(0)}$)~\cite{Beh1982}, where $K=3$ denotes the degree of forbiddenness. The second approach uses each model's standard value (SV) for the sNME. The results of these calculations are presented in Table~\ref{Table:HalfLives}.

Remarkably, all the four models produce very similar values for the $\ell$-NME and the CVC and standard values for the sNME, as the table shows. Furthermore,
neither methodology reproduces the experimental half-life of $T_{1/2} = 5.08(13) \times 10^{10}$~yr, irrespective of the chosen $g_{\rm A}$  value. This discrepancy is expected; the NSM operates within a limited valence space, whereas the sNME also receives contributions from configurations outside this space. Consequently, the sNME computed by the models (the SV for it) is inherently incomplete, while the CVC-derived value represents an idealized theoretical estimate. Therefore, we next employ the ESSM methodology, as described in the theoretical background section, to simultaneously reproduce the experimental half-life and determine the corresponding theoretical $\beta$-spectral shapes.

\begin{table*}[hbtp]
\centering
\caption{Calculated half-lives for the implemented models as a function of the axial-vector coupling constant, $g_{\rm A}$, with the sNME as the CVC-value or the standard value (SV). For completeness, the CVC and SV values are listed alongside the $\ell$-NME, for each model. All computations conserve $g_{V}=1$.}
\resizebox{\textwidth}{!}{%
\begin{tabular}{c c c c c c c c c c c c c}
\hline\hline
& \multicolumn{3}{c}{\textbf{Nuclear Matrix Elements}} & & & \multicolumn{7}{c}{\textbf{Computed Half-Life} ($\times 10^{10}$ years)} \\
\cmidrule(lr){2-4} \cmidrule(l){7-13}
\small\textit{Interaction} & \small\textit{CVC} & \small\textit{\(\ell\)-NME} & \small\textit{SV} & \small\textit{Method} & $g_{\rm A}$: 
  & \small\textit{0.20}
  & \small\textit{0.40}
  & \small\textit{0.60}
  & \small\textit{0.80}
  & \small\textit{1.00}
  & \small\textit{1.20}
  & \small\textit{1.40} \\
\hline
\multirow{2}{*}{MQPM} & \multirow{2}{*}{-3.13} & \multirow{2}{*}{-191} & \multirow{2}{*}{1.82} & CVC & & 0.208 & 0.165 & 0.133 & 0.109 & 0.091 & 0.077 & 0.066 \\
 &  &  &  & SV & & 0.101 & 0.123 & 0.153 & 0.196 & 0.259 & 0.354 & 0.506 \\
\hline
\multirow{2}{*}{glbepn} & \multirow{2}{*}{-3.50} & \multirow{2}{*}{-214} & \multirow{2}{*}{1.75} & CVC & & 0.175 & 0.145 & 0.121 & 0.102 & 0.087 & 0.075 & 0.065 \\
 &  &  &  & SV & & 0.092 & 0.109 & 0.132 & 0.162 & 0.204 & 0.263 & 0.349 \\
\hline
\multirow{2}{*}{jj44b} & \multirow{2}{*}{-2.82} & \multirow{2}{*}{-172} & \multirow{2}{*}{1.65} & CVC & & 0.260 & 0.207 & 0.168 & 0.139 & 0.116 & 0.098 & 0.084 \\
 &  &  &  & SV & & 0.122 & 0.148 & 0.183 & 0.231 & 0.300 & 0.402 & 0.561 \\
\hline
\multirow{2}{*}{jun45} & \multirow{2}{*}{-2.94} & \multirow{2}{*}{-179} & \multirow{2}{*}{1.72} & CVC & & 0.240 & 0.193 & 0.158 & 0.131 & 0.110 & 0.093 & 0.080 \\
 &  &  &  & SV & & 0.112 & 0.135 & 0.165 & 0.206 & 0.264 & 0.348 & 0.476 \\
\hline\hline
\end{tabular}%
 }
\label{Table:HalfLives}
\end{table*}

\begin{figure}[!ht]
\centering
\includegraphics[width=\textwidth]{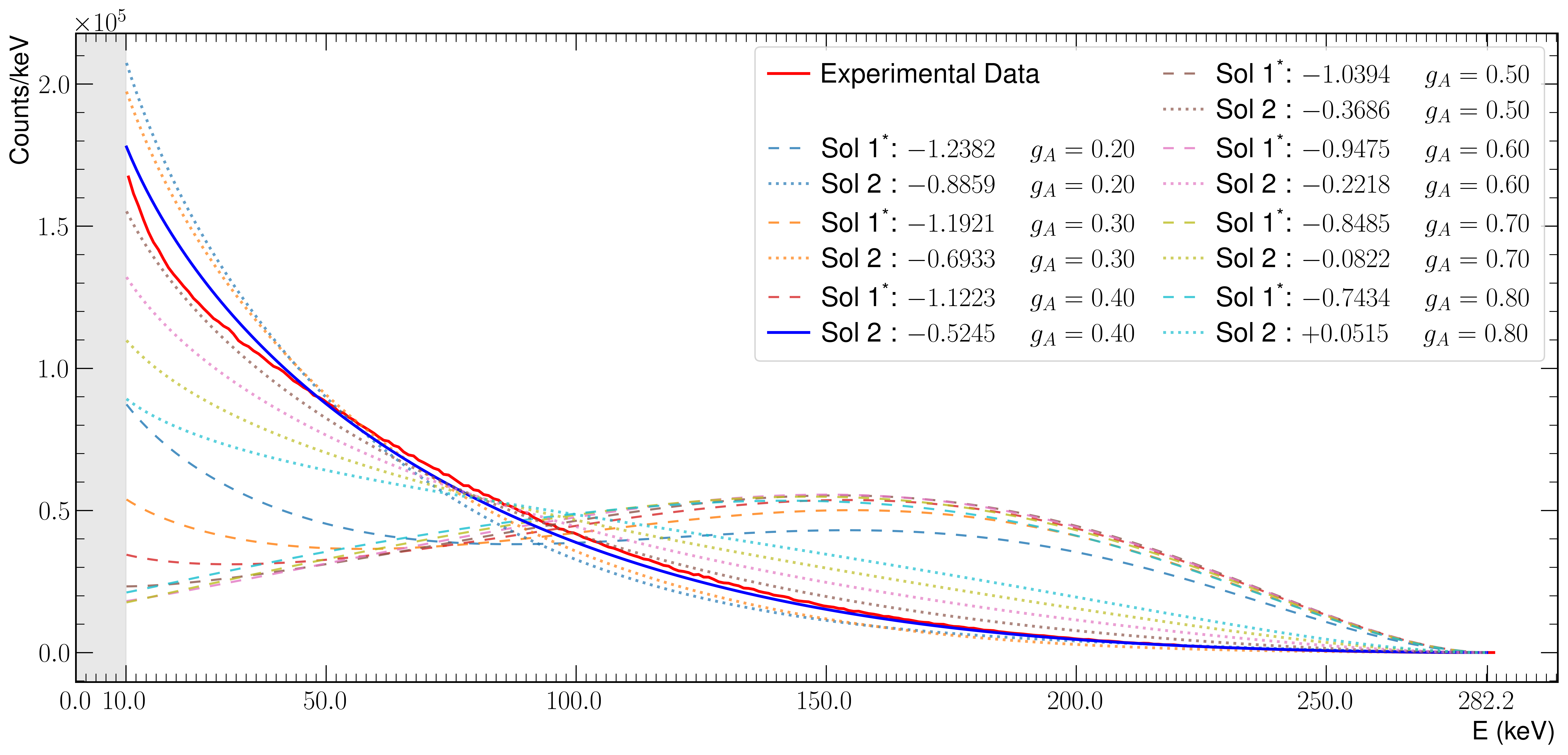}
\caption{The experimental unfolded $\beta$ spectrum of $^{87}\text{Rb}$ (solid red line, this work) is compared with theoretical ESSM predictions from the shell model using the \texttt{jun45} interaction. The theoretical spectra have been normalized to unity and scaled to total experimental counts. The half-life-reproducing sNME pairs (dashed/dotted lines) are obtained at discrete values of the $g_{\rm A}$ when varied from $0.2-0.8$. The best theoretical match to the data is presented as the solid blue line. Solutions marked with an asterisk denote sNME(c) (closer to the CVC value), and unmarked ones, sNME(f) (farther from the CVC value).}
 \label{jun45_comparison}
\end{figure}

The theoretical $\beta$-spectral shapes were generated following the ESSM methodology with the effective axial-vector coupling ($g_{\rm A}$) varied from 0.2 to 0.8 in increments of 0.1. (The explorations had initially extended up to $g_{\rm A} = 1.27$, encompassing the commonly accepted range for this parameter.) For each selected $g_{\rm A}$ value, the small Nuclear Matrix Element (sNME) was adjusted until the calculated half-life reproduced the experimental value of $T_{1/2} = 5.08(13) \times 10^{10}$~yr for $^{87}\text{Rb}$. Due to the quadratic dependence of the integrated shape function $C(W_e)$ on the sNME, up to two sNME solutions typically exist that satisfy this half-life constraint for a given $g_{\rm A}$. This procedure yields pairs of ($g_{\rm A}$, sNME) values, though in some instances, no real sNME solution may be found if the underlying nuclear model and chosen $g_{\rm A}$ cannot reproduce the experimental 
half-life. These solution pairs are illustrated for the \texttt{jun45} interaction in Fig.~\ref{jun45_comparison}.

Figure~\ref{jun45_comparison} displays the experimentally unfolded $\beta$-spectrum obtained in this work alongside the several theoretical spectral shapes in the ESSM scheme generated from the ($g_{\rm A}$, sNME) solution pairs for the \texttt{jun45} interaction. To facilitate a direct comparison where the experimental spectrum's magnitude depends on factors such as measurement duration, sample activity, and detector efficiency, our theoretical spectra were first normalized to unity (i.e., unit area under the curve), then these normalized spectra were scaled by the total number of counts in the experimental spectrum, ensuring independence from the number of atoms to decay in the side-by-side analysis, then a reduced chi-square ($\chi^2_\nu$) analysis was performed for each scaled theoretical spectrum against the experimental data to identify the optimal ($g_{\rm A}$, sNME) pair yielding the best agreement. The analysis utilized 1~keV wide energy bins. Given the $Q$-value of 282.275(6)~keV for the $^{87}\text{Rb}$ decay, this binning results in 271 (number of bins minus the low energy threshold and the one free parameter) degrees of freedom for the $\chi^2_\nu$ calculation. The theoretical spectrum providing the minimum $\chi^2_\nu$ is considered the best match. For this comparison and the $\chi^2_\nu$ analysis, a low-energy threshold of 10~keV was applied, corresponding to the gray-shaded region in Fig.~\ref{jun45_comparison}. 
The best $\chi^2$ is obtained by the 'Sol~2' solution with sNME~=~-0.5245 at $g_{\rm A} = 0.40$, depicted by the solid blue line 
in the plot for the \texttt{jun45} interaction.

This ESSM analytical procedure, involving the determination of ($g_{\rm A}$, sNME) pairs by fitting to the experimental half-life and subsequent $\chi^2_\nu$ minimization against the experimental $\beta$ spectrum, was identically applied to the remaining shell-model interactions (\texttt{glbepn}, \texttt{jj44b}) and the \texttt{MQPM}. The three best-fit solutions (i.e., those yielding the lowest $\chi^2_\nu$ values) for each model are summarized in Table~\ref{tab:chi_square_table}. As evident from the table, optimal agreement is generally achieved for $g_{\rm A}$ values in the range of 0.4 to 0.6 (the latter specifically for \texttt{glbepn}). The second and third best-fit solutions for each model consistently exhibit progressively larger $\chi^2_\nu$ values. Although Table~\ref{tab:chi_square_table} presents only these top three solutions per model for conciseness, the $\chi^2_\nu$ values typically increase sharply for $g_{\rm A}$ values outside this preferred 0.4--0.6 range, further reinforcing the optimal region for $g_{\rm A}$. It should be noted that the relatively large magnitudes of the $\chi^2_\nu$ values, even for the best-fit solutions, are a consequence of the high statistical precision of the experimental data, with counts per bin reaching the order of $10^5$, the corresponding statistical uncertainty, $\sigma=N$, becomes exceedingly small relative to the signal in each bin. The $\chi^2_\nu$ test quantifies the disagreement between the model and data in units of this small uncertainty. Consequently, the analysis becomes a highly stringent test where visually imperceptible differences between the theoretical shape and the experimental spectrum are revealed to be statistically significant deviations of many $\sigma$, highlighting even subtle differences between the models and the data.

Regarding the two sNME solution branches obtained for each $g_{\rm A}$---denoted sNME(c) (closer to the Conserved Vector Current (CVC) hypothesis prediction) and sNME(f) (farther from the CVC prediction)---a distinct preference emerged across all models. Specifically, only the sNME(f) solutions yielded $\beta$-spectral shapes with $\chi^2_\nu$ values indicative of a reasonable agreement with the experimental data. This outcome is consistent with our findings from several previous studies \cite{Kos2023,Kos2024,Ram2024,Ram2024b,Ram2024c}, where no systematic preference for sNME solutions adhering to the CVC-based value was observed. This is attributed to the fact that the CVC hypothesis, which relates the sNME to the leading-order NME ($\ell$-NME), strictly applies within the framework of a perfect many-body theory. Such ideal conditions are not fully met by nuclear models, which operate with truncated valence spaces and assume a closed core. Under the CVC hypothesis, the sNME is expected to be proportional to the $\ell$-NME. While the calculated $\ell$-NMEs differ for each interaction employed in this work (as shown in Table~\ref{Table:HalfLives}, the sNME(f) values presented in Table~\ref{tab:chi_square_table} for a given $g_{\rm A}$ are notably similar across the different models, 
possibly due to the dominating rank-3 and rank-4 one-body transition density from $0g_{9/2} \rightarrow 1p_{3/2}$, with similar magnitude across all models. Lastly, a side-by-side comparison of the CVC values in Table~\ref{Table:HalfLives} and the sNME(f) values in Table~\ref{tab:chi_square_table} confirms that the CVC value alone is insufficient to reproduce the experimental data. Furthermore, the ratio between the CVC and the required sNME(f) is large, ranging from $5.75$ for MQPM to $6.82$ for glbepn.

\begin{table}[htbp]
\centering
\footnotesize
\setlength{\tabcolsep}{2.5pt}
\caption{Reduced chi-square ($\chi^2_\nu$) values from the comparison of theoretical spectral shapes with the unfolded experimental $^{87}\text{Rb}$ data for each model are presented. The parameters defining the theoretical shapes ($g_{\rm A}$, sNME) were determined as detailed for Fig.~\ref{jun45_comparison}. The best overall matches per model were found within the sNME(f) solution set. Each combination yields half-lives of $5.07644(12)\times 10^{10}$ (by design) using the solution set sNMEs regardless of their ($\chi^2_\nu$) values.}
\label{tab:chi_square_table}
\begin{tabular*}{\textwidth}{l @{\extracolsep{\fill}} ccc ccc ccc @{}}
\toprule
\textbf{Model} & \textbf{$g_{\rm A}^{\text{(f)}}$} & \textbf{sNME} & \textbf{$\chi^2_\nu$} & \textbf{$g_{\rm A}^{\text{(f)}}$} & \textbf{sNME} & \textbf{$\chi^2_\nu$} & \textbf{$g_{\rm A}^{\text{(f)}}$} & \textbf{sNME} & \textbf{$\chi^2_\nu$} \\
\cmidrule(lr){2-4} \cmidrule(lr){5-7} \cmidrule(lr){8-10}
\midrule
\texttt{jj44b}  & 0.30 & -0.621 & 621 & 0.40 & -0.459 & 100 & 0.50 & -0.308 & 1584 \\
\texttt{glbepn} & 0.50 & -0.687 & 781 & 0.60 & -0.514 & 116 & 0.70 & -0.354 & 1180 \\
\texttt{jun45}  & 0.30 & -0.693 & 992 & 0.40 & -0.524 & 147 & 0.50 & -0.369 & 705 \\
\texttt{MQPM}   & 0.30 & -0.740 & 1130 & 0.40 & -0.544 & 190 & 0.50 & -0.365 & 827 \\
\bottomrule
\end{tabular*}
\end{table}

\subsection{Analysis on the weak-axial coupling obtained}
\begin{figure}[!htp]
\centering
\includegraphics[width=\textwidth]{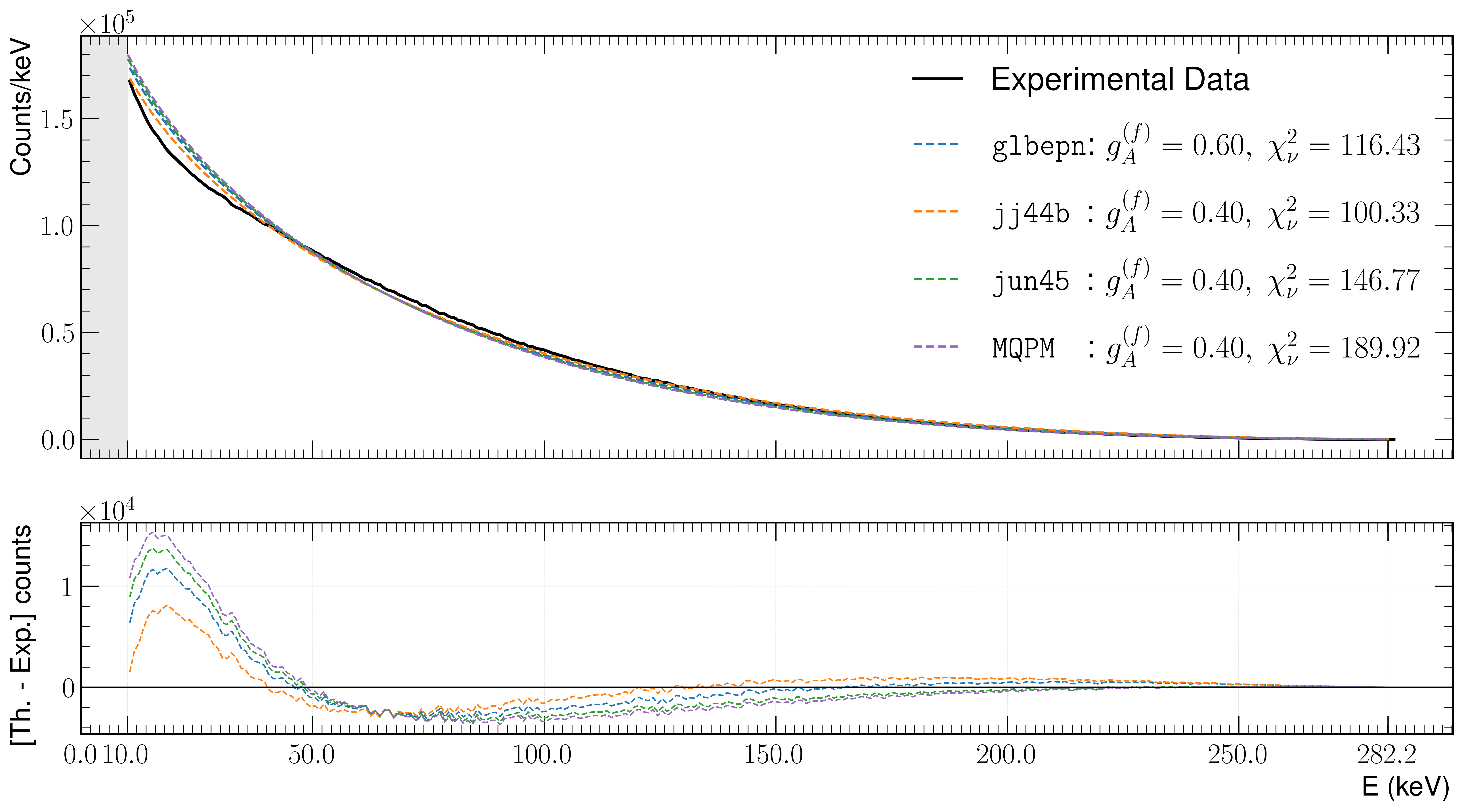}
\caption{Comparison of the experimental unfolded $\beta$ spectrum (solid black line) with the best-match ESSM spectral shapes. These include shell-model calculations with the \texttt{glbepn}, \texttt{jj44b}, and \texttt{jun45} interactions, and predictions from the Microscopic Quasiparticle-Phonon Model (\texttt{MQPM}). For each model, the displayed optimal fit to the experimental data is determined through minimization of the reduced chi-square, following the methodology detailed for Fig~\ref{jun45_comparison}. The legend specifies the $g_{\rm A}^{(f)}$ value (corresponding to the sNME(f) solution branch) and the associated $\chi^2_\nu$ statistic for each presented spectrum. The lower panel displays the corresponding residual counts. (Color online)}
 \label{curves_comparison}
\end{figure}

A consolidated comparison of the best-fit theoretical $\beta$-spectral shapes from each model (shell-model interactions \texttt{glbepn}, \texttt{jj44b}, \texttt{jun45}, and the \texttt{MQPM}) with the experimentally unfolded data is presented in Fig.~\ref{curves_comparison}. Remarkably good agreement between theory and experiment is achieved for the optimal $g_{\rm A}$ values identified for each model. By construction, these best-fit theoretical spectra simultaneously reproduce the experimental half-life (used as a constraint in determining the sNME) and provide the closest match to the $\beta$-spectral shape. This simultaneous reproduction is a significant outcome, as many theoretical approaches do not achieve it. The methodology employed here, which uses the half-life solely as a constraint to determine the sNME for a given $g_{\rm A}$ before comparing the spectral shape, ensures a robust analysis that is not merely tuned to fit the spectral data directly without this physical constraint. Employing the experimental half-life as a primary constraint is a sound methodological choice, given that this quantity is often well-measured; potential uncertainties in branching ratios, which can be a concern in other decay scenarios, are less critical when a single decay branch dominates or is precisely known.

The theoretical calculations incorporated standard radiative and screening corrections. Additionally, an atomic exchange correction 
must be added. A general framework concerning this correction for forbidden non-unique beta decays was displayed in \cite{Moug2023} and then simplified to a practically applicable correction for forbidden unique beta decays in the same work. Since a practical way to implement the exchange correction for non-unique beta decays is still lacking, we adopt the exchange correction recently developed for allowed decays in \cite{Nitescu2023} as a surrogate in the present work.
This specific exchange correction \cite{Nitescu2023} is known to enhance the electron energy distribution at low energies, typically in the $0-50$~keV range. Consequently, its application contributes to the consistent overestimation of the $\beta$-spectrum intensity by all theoretical models applied herein in this low-energy region when compared to the experimental data, as can be observed in Fig.~\ref{curves_comparison}. This discrepancy in the low-energy region is a primary contributor to the resulting $\chi^2_\nu$ values being significantly greater than unity (i.e., $\chi^2_\nu \gg 1$). Nevertheless, despite their elevated absolute magnitudes, these $\chi^2_\nu$ values remain a valid and useful metric for ranking the relative agreement of the various theoretical curves with the experimental data, as this introduces a consistent bias propagated equally 
throughout the models in this work.

The optimal effective axial-vector coupling constant ($g_{\rm A}$) derived from fitting the experimental data in the present study indicates a significantly stronger quenching than is typically observed. Most recent phenomenological studies report $g_{\rm A}$ values in the range of $0.7-1.1$, including recent investigations such as those detailed in Refs.~\cite{Kos2023,Kos2024}. This pronounced deviation from typical $g_{\rm A}$ values necessitated a series of rigorous checks of the analysis procedure and underlying assumptions. These investigations included, but were not limited to 
an assessment of the sensitivity of the results to the adopted $Q$-value and half-life (282.276(6) keV and $5.08(13)\times 10^{10}$ yr) by varying the central values of these quantities independently and simultaneously by $\pm 5\%$, which resulted in only minor adjustments, about $\Delta g_{\rm A}\approx\pm 0.1$. In addition, the above-cited experimental uncertainties
were propagated via an envelope analysis for the \texttt{jj44b} interaction. The envelope consists of two extreme scenarios, corresponding to the slowest and fastest possible decay rates within the $1\sigma$ limits. The analysis confirmed a stable value of $g_{\rm A} = 0.40$, with the sNME constrained to $-0.4592^{+0.0004}_{-0.0001}$ and a corresponding $\chi^2_\nu$ of $100.3^{+15.5}_{-11.4}$. The stability of $g_{\rm A}$ across this range demonstrates that its determination is robust and insensitive to the uncertainties in the input experimental data; III) enhancements to the $\beta$-decay calculation codes to incorporate more realistic nuclear radii derived from experimental charge radii, accounting for the finite size of the proton (with comparative calculations also performed using the conventional $R = r_0 A^{1/3}$ approximation); and IV) exploration of the sensitivity to the \texttt{glbepn} model space definition, where expanding a more truncated initial space (comparable to that of \texttt{jj44b} and \texttt{jun45}) to allow for greater configuration mixing yielded similar optimal $g_{\rm A}$ values. On the other hand, in the present case, expanding the model space does  likely not help since there is one dominating single-particle transition within the present model space and another outside this model space is not likely to be as strong. Furthermore, the MQPM calculation supports this argument since also it features the same one dominating single-particle transition although the model space of MQPM is very large.

Ultimately, it is hypothesized that this unusually strong quenching primarily stems from the specific nuclear structure involved: the decay of the semi-magic $^{87}\text{Rb}$ ($Z=37, N=50$) to $^{87}\text{Sr}$ ($Z=38, N=49$), which involves creating a neutron hole relative to the magic $N=50$ shell closure. Given that the effective shell-model interactions employed are typically optimized to reproduce properties across a broader region of nuclei, rather than being fine-tuned for individual isotopes or highly specific shell configurations, their ability to precisely capture the nuances of such a transition could be limited. This could lead to an overestimation of the wavefunctions' overlap and thus deviations from the `expected' $g_{\rm A}$ range of $0.7-1.1$ for all interactions tested. We have conducted preliminary tests that lend some support to this hypothesis. For instance, calculations for the allowed decay of $^{85}\text{Br}$ (another nucleus with $N=50$) using the \texttt{jj44b} interaction required a $g_{\rm A}\approx 0.47$ to reproduce its experimental half-life, indicating similar strong quenching. In contrast, the allowed decay of $^{80}\text{Br}$, further from this specific shell closure, required $g_{\rm A}\approx 0.97$ with the same interaction, a value more consistent with typical quenching factors. Although these are preliminary findings, they suggest that the specific shell structure and regional fitting of the interactions may be key factors. A more systematic investigation across a wider range of isotopes with well-established experimental data is necessary to conclusively validate or refute this hypothesis, eventually shedding light on the real explanation at hand.

\section{Conclusions}

In this paper we present experimental details of the measurement and theoretical analysis of the $\beta$ spectral shape of electrons emitted in the third-forbidden non-unique ground-state-to-ground-state $\beta$-decay transition $^{87}\textrm{Rb}(3/2^-)\to\,^{87}\textrm{Sr}(9/2^+)$, with a 100\% branching ratio
having the  aim to determine the effective value of the weak axial coupling $g_{\rm A}$.
In particular, in order to effectively measure the $^{87}$Rb $\beta$-decay, a new Rb$_2$ZrCl$_6$ crystal scintillator was realized assuring 
the possibility to exploit the  effective ``source=detector" approach, where the source is embedded in the constituting materials of the detector itself.  
The production and the characterization of this crystal have been studied and described in detail.  The data taking was performed deep underground at 
the Gran Sasso National Laboratory of INFN. The experimental $\beta$ spectrum was measured and uncertainties at low energies were addressed.  
The experimental half-life of the process was measured as T$_{1/2} = 5.08(13) \times$ 10$^{10}$ yr, 
which is -- within the uncertainties -- in agreement with previous 
measurements that can be found in a recent decay data evaluation \cite{DDEP2025}. 
The uncertainty of this measurement is primarily driven by the red band in Fig. \ref{spect-band}, which could be reduced by addressing the overshoot issue through appropriate hardware modifications. Implementing such improvements could also lower the energy threshold, representing a significant step toward a more precise determination of the half-life.
Moreover, the determined $\beta$ spectral shape was also analysed in order to try to pin down the effective value of the weak axial coupling $g_{\rm A}$, which can play a relevant role for the sensitivity estimations of present and future experiments beyond the standard model
\cite{Eng2017,Eji2019,Ago2023}. With this purpose the data have been investigated within several theoretical models driving to a quenched value of the axial coupling constant, $g_{\rm A}$ in the range 0.4--0.6. In particular, we fit the value of the so-called small relativistic nuclear matrix element in order to reproduce the measured half-life of the $\beta$ transition in question;  this phenomenological method was named Enhanced Spectrum-Shape Method (ESSM).   
Related discussions were addressed. As for the perspectives, in the future the theoretical approach could benefit from implementation of relativistic electron wave functions obtained directly as solutions of the Dirac equation. Presently, we use only analytical correspondences of these wave functions, thus losing precision at short radial distances. Also radiative corrections could be improved, although their contributions are minor in these calculations. 
One very relevant improvement would be an implementation of practically applicable atomic exchange corrections for forbidden non-unique beta transitions, deduced from the corresponding general framework displayed in \cite{Moug2023}.
This would greatly improve the problematic situation of the present comparison at low energy.
In the future, we could also expect new improved theoretical works and experimental measurements to assess this relevant topic. In particular, in near future new experimental measurements are foreseen by various activities using various isotopes and considering the promising method of comparison between computed and measured electron spectra of high-forbidden non-unique $\beta$-decays.

\end{document}